\documentclass{pasj00}

\begin{document}
\SetRunningHead{Imanishi \& Nakanishi}{IR spectroscopy and HCN/HCO+
emission from Arp 299} 
\Received{2006/6/7}
\Accepted{2006/8/14}

\title{Infrared 2--4 Micron Spectroscopy and Millimeter
Interferometric HCN and HCO$^{+}$ Observations of the
Individual Merging Components of Arp299} 

\author{Masatoshi Imanishi} 
\affil{National Astronomical Observatory, 2-21-1, Osawa, Mitaka, Tokyo
181-8588, Japan} 
\email{masa.imanishi@nao.ac.jp} 

\and

\author{Kouichiro Nakanishi}
\affil{Nobeyama Radio Observatory, Minamimaki, Minamisaku, Nagano,
384-1305, Japan} 
\email{nakanisi@nro.nao.ac.jp}

\KeyWords{galaxies: active --- galaxies: nuclei --- galaxies: ISM --- 
radio lines: galaxies --- infrared: galaxies --- galaxies: individual
(Arp299)} 

\maketitle

\begin{abstract}
We present ground-based infrared $K$- (2--2.5 $\mu$m) and $L$-band
(2.8--4.1 $\mu$m) spectroscopy, as well as interferometric
observations at $\lambda \sim$ 3mm, for the individual merging
components (A, B, and C) of the luminous infrared galaxy Arp 299. We
investigate the presence and location of the putative buried active galactic nucleus (AGN) inferred from previous X-ray observations at E $>$ 10 keV.
Our sub-arcsec-resolution infrared spectra clearly
reveal that the putative buried AGN resides in the nucleus B1 (a
subcomponent of B), based on a very low equivalent width of 3.3
$\mu$m polycyclic aromatic hydrocarbon emission, a weak 2.3 $\mu$m
CO absorption feature, and a large time variation of the $K$- and
$L$-band continuum fluxes. In component C, we find strong 3.1 $\mu$m ice
absorption at $L$ and weak 2.3 $\mu$m CO absorption at $K$,
as expected in a buried AGN; however, a centrally concentrated young
super star cluster is an alternative possibility because of the
modest infrared luminosity and non-galaxy-nucleus nature of this
component. The infrared $K$- and $L$-band spectra of the infrared
brightest nucleus, A, are typical of a normal starburst with no
explicit AGN signatures. Our interferometric observations
simultaneously obtain HCN (J = 1--0) and HCO$^{+}$ (J = 1--0) emission
lines with $\sim$4 arcsec resolution, and we find the HCN to HCO$^{+}$
brightness-temperature ratios to be as low as those found in starburst
nuclei in all the major merging components of Arp 299. 
The low ratio even in the AGN-hosting nucleus B may be due
to the presence of a large amount of high-density molecular gas whose
chemistry is dominated by coexisting starbursts and/or shocks, rather
than by the central strong X-ray-emitting AGN. 
\end{abstract}

\section{Introduction}

Luminous infrared galaxies (LIRGs; L$_{\rm IR}$ $\gtrsim$
10$^{11}$L$_{\odot}$), discovered by the {\it IRAS} all sky survey,
radiate the bulk of their luminosities as dust emission in the
infrared \citep{sam96}. Powerful energy sources, starbursts and/or
active galactic nuclei (AGNs), must therefore be present, but hidden
behind dust. Since these LIRGs dominate the cosmic infrared background
emission, which is the sum of dust-obscured activity in the universe
\citep{dol06}, distinguishing their energy sources is closely related
to clarifying the connections between obscured AGN and starbursts.

The difficulty in disentangling an AGN from a starburst as the primary
energy source of a LIRG comes from the fact that highly concentrated
molecular gas and dust in a LIRG's nucleus \citep{sam96} can easily
{\it bury} an AGN in all directions. A luminous AGN surrounded by
torus-shaped dust is distinguishable from a normal starburst galaxy
relatively easily through optical and infrared
high-resolution spectroscopy \citep{vei87,gen98} because emission
from the so-called narrow line regions (NLRs), which are
well developed along the torus axis above a torus scale height and
are photoionized by the central AGN's radiation, is visible from all
directions \citep{ant93}. However, the presence of a {\it buried} AGN
in a LIRG's core is difficult to unveil with this conventional type of
spectroscopy because no significant NLRs are expected to develop.

Even if a luminous buried AGN is present in a LIRG's core, starbursts
are likely to coexist, surrounding the central AGN. When such a
starburst/buried-AGN composite LIRG is observed at wavelengths of
large dust extinction, the observed flux is dominated by the less
obscured starbursts, with a small contribution from the more highly
obscured buried AGN. It is therefore essential to observe at
wavelengths of low dust extinction if we are to properly investigate 
the presence of a buried AGN in a LIRG's core. Several useful methods
for this purpose currently exist. The first is to observe in 
the hard X-ray band at E $>$ 2 keV. Since an AGN is intrinsically a
much stronger X-ray emitter than a starburst \citep{elv94,ran03}, the
detection of strong but highly-absorbed X-ray emission can provide
evidence of a luminous buried AGN in a LIRG's core. The second is
millimeter interferometric searches for the chemical effects of the
strong X-ray-emitting AGN on the surrounding interstellar medium.
Around a buried AGN, X-ray dissociation regions (XDRs; Maloney et al.
1996) should develop, rather than the photodissociation regions
(PDRs) usually found in starburst regions. Since the emission
mechanism is different, XDRs and PDRs can be distinguished if they
show different flux ratios of emission lines. The feasibility of
carrying out this distinction has been demonstrated using HCN (J = 1--0)
($\lambda$ = 3.3848 mm) and HCO$^{+}$ (J = 1--0) ($\lambda$ = 3.3637 mm)
emission lines in the millimeter wavelength range \citep{koh05}. The
third method is infrared spectroscopy. While a normal starburst should
always show large-equivalent-width polycyclic aromatic hydrocarbon
(PAH) emission features in the spectra at $\lambda_{\rm rest}$ = 3--25
$\mu$m in the rest-frame, an AGN displays a PAH-free continuum
\citep{gen98,imd00}. Infrared $K$-band (2.0--2.5 $\mu$m) spectroscopy
can also be used to separate starburst and AGN emission, because a
normal starburst displays CO absorption features at $\lambda_{\rm rest}$
= 2.3--2.4 $\mu$m, whereas an AGN does not \citep{iva00,ia04}. Since a
buried AGN is generally more difficult to find with any method than the
surrounding less obscured starbursts, using a combination of these
potentially powerful approaches can increase the likelihood of detecting
the elusive buried AGN and help to draw a quantitatively consistent
picture about the energetic importance of the buried AGN.

Arp 299 ($z$ = 0.010) is a nearby well studied LIRG. It consists of
three main merging components, the eastern component IC 694 (Arp 299
A) and the western components making up NGC 3690 (Arp 299 B and C)
\citep{geh83,wyn91}. The Arp 299 B nucleus is divided into
subcomponents (B1 and B2) with a separation of $\sim$2.5 arcsec in
the infrared \citep{wyn91,zho93,mil96,ket97,lai99,soi01}. The Arp 299
C region is taken to be a ``non-nucleus''\citep{dud93,cha02} and
consists of the C and C' components with a separation of $\sim$5 arcsec in
the infrared \citep{zho93,mil96,ket97,lai99,soi01}. A component D is
detected in the radio at 6 cm (5 GHz) \citep{nef04}, but is not
found in the deepest infrared image \citep{soi01}. The whole
merging system has an infrared luminosity of L$_{\rm IR}$ $\sim$ 5
$\times$ 10$^{11}$L$_{\odot}$ (= 10$^{45.3}$ ergs s$^{-1}$; see Table
1), approximately 60\% of which comes from the A (IC 694)
nucleus, based on observations using the {\it Kuiper Airborne
Observatory} \citep{joy89,cha02}. Although no explicit AGN signatures
have been found in any nucleus through optical spectroscopy
\citep{bus84,kee85,arm89,coz98}, highly-absorbed (Compton-thick;
N$_{\rm H}$ $>$ 10$^{24}$ cm$^{-2}$) but intrinsically strong X-ray
emission has been detected at E $>$ 10 keV \citep{del02}, suggesting the
presence of a luminous buried AGN. Unfortunately, because of the lack
of spatial resolution of the X-ray data at E $>$ 10 keV, it is unclear
which nucleus (or nuclei) possesses the buried AGN. While the
equivalent width of the 6.4-keV iron K$\alpha$ line (EW$_{K \alpha}$)
in higher spatial-resolution X-ray data at E $<$ 10 keV was used to
argue that a putative AGN resides in nucleus B1, at least
\citep{zez03,bal04}, it is unclear whether an AGN is present in any of
the other nuclei because the 6.4-keV emission can be weak in a
Compton-thick {\it buried} AGN \citep{fab02}. Further observational
data are required to better understand the role of the putative
buried AGN in Arp 299.

To spatially separate the subcomponents of the Arp 299 merging
system, observations with spatial resolution better than a few to 5
arcsec are desirable. Ground-based infrared $K$- (2.0--2.5 $\mu$m) and
$L$-band (2.8--4.1 $\mu$m) spectroscopy using a large ($>$3m)
telescope at a good observing site can achieve sub-arcsec spatial
resolution and so is suitable for this purpose. At millimeter
wavelengths, interferometry is better suited than a single-dish radio
telescope to obtain morphological information, by resolving
the subcomponents, and to investigate the properties of molecular gas at
nuclear regions where AGNs are thought to exist, by reducing the
contamination from extended emission. 
In this paper, we report on the results of sub-arcsec-resolution
infrared $K$- and $L$-band spectroscopy and millimeter interferometric
maps with $\sim$4 arcsec 
resolution, of the HCN (J = 1--0) and HCO$^{+}$ (J = 1--0) emission lines
for the individual main merging components of Arp 299.
The aim is to identify the location(s) of the putative buried AGN. 
Throughout this paper, $H_{0}$ $ = $ 75 km s$^{-1}$ Mpc$^{-1}$,
$\Omega_{\rm M}$ = 0.3, and $\Omega_{\rm \Lambda}$ = 0.7 are adopted, so
that the distance to Arp 299 is 40.3 Mpc and 1 arcsec corresponds to 190
pc. 

\section{Observations and Data Reduction}

\subsection{Infrared $K$- and $L$-band Spectroscopy} 

Infrared $K$- (2.0--2.5 $\mu$m) and $L$-band (2.8--4.1 $\mu$m) spectra
of the Arp 299 A (IC 694), B and C were taken using SpeX \citep{ray03}
attached to the IRTF 3-m telescope atop Mauna Kea, Hawaii, on 2005
April 27 (UT). The 1.9--4.2 $\mu$m cross-dispersed mode was employed,
so that $K$- and $L$-band spectra were obtained simultaneously. For
Arp 299, narrowband imaging data centered on the 3.3 $\mu$m PAH
emission feature were available \citep{sat99} and revealed that the
3.3-$\mu$m PAH emission is spatially extended, with a size of 2 to 3
arcsec. The imaging data can provide a rough absolute observed
3.3-$\mu$m PAH-emission luminosity and its ratio relative to the
infrared luminosity \citep{sat99}. We therefore chose a relatively
narrow 0\farcs8 wide slit, to pinpoint the continuum emission peak of
each merging component where the putative AGN may be located.
Accordingly, the extended starburst emission is not fully covered with
our infrared spectra. We search for buried AGN signatures at the
nucleus, based on the {\it equivalent width} of emission and
absorption features in the $K$- and $L$-bands, since these
equivalent width values cannot be obtained in detail from the
narrowband imaging data \citep{sat99}. The resulting spectral
resolution using this slit is R $\sim$ 1000 in the $K$- and
$L$-bands.

Observations of Arp 299 A, B (B1+B2), and C were performed
independently. The position angle of the slit was set along the
north--south direction for Arp 299 A and C, while for Arp 299 B it was
37$^{\circ}$ west-of-north to accommodate the signals from the B1 and
B2 nuclei simultaneously \citep{zho93,sat99,lai99}. Spectroscopy of
Arp 299 C$'$ was not attempted because its $K$- and $L$-band fluxes
reported in the literature \citep{zho93} are more than 1 mag fainter
than the other components, and thus it would be difficult to obtain a
meaningful spectrum within the limited observing time available to us.
The total net on-source integration times were 60, 60, and 30 min for
components A, B (B1+B2), and C, respectively. The sky conditions
were photometric throughout the observations. The seeing at $K$ was
measured to be in the range 0$\farcs$8--1$\farcs$0 in full-width at
half-maximum (FWHM). A standard telescope nodding technique (ABBA
pattern) with a throw of 7$\farcs$5 was employed along the slit. The
telescope tracking was monitored with the infrared slit-viewer of
SpeX. Each exposure was 15 sec, and 2 coadds were made at each
position.

The F8V main sequence star HR 4112 (V = 4.84) was observed as a standard
star, with an airmass difference of $<$0.1 to the individual main merging
components, to correct for the transmission of the Earth's atmosphere.
The magnitude of the standard star HR 4112 was estimated to be $K$ =
3.49 and $L$ = 3.45, based on its $V$-band (0.6 $\mu$m) magnitude (V =
4.84) and the color of an F8V star ($V-K$ = 1.35 and $V-L$ = 1.39)
\citep{tok00}. For reference, the $K$-band magnitude of HR 4112 in the
2MASS database \citep{skr06} is $K$ = 3.64$\pm$0.22 mag, consistent with
the above estimated value within the uncertainty. Standard data
reduction procedures were employed using IRAF
\footnote{
IRAF is distributed by the National Optical Astronomy Observatories,
operated by the Association of Universities for Research
in Astronomy, Inc. (AURA), under cooperative agreement with the
National Science Foundation.}. 
Initially, bad pixels and pixels hit by cosmic rays were replaced with
values interpolated from the surrounding pixels. Then, frames taken
with an A (or B) beam were subtracted from frames subsequently taken
with a B (or A) beam, and the resulting subtracted frames were added,
then divided by a spectroscopic flat image. The spectral extraction of
Arp 299 merging components along the slit was then made by integrating
signals over 3$\farcs$0 (0.15 arcsec pixel scale $\times$ 20 pixels)
because the emission is slightly spatially extended. For HR 4112, the
actual image size along the slit tended to increase with decreasing
wavelength, and was different for the different datasets taken at
different times, due to changes in the seeing size. Thus, the pixel
number used for the spectral extraction of HR 4112 varied, but was
usually smaller than the value of 3$\farcs$0 employed for Arp 299.
Wavelength calibration was performed using the wavelength-dependent
transmission of the Earth's atmosphere. The spectra of Arp 299's
individual merging components were divided by that of the standard
star HR 4112, and were multiplied by a blackbody spectrum with a
temperature of 6000 K, corresponding to an F8V star \citep{tok00}.
Appropriate spectral binning was applied to reduce scatter. Flux
calibration was made using signals detected inside our slit spectra.
Since the employed slit width was comparable to or slightly smaller
than the seeing size, our slit spectra may have introduced
a small amount of flux ambiguity even for the compact (spatially
unresolved) emission at the peak of each merging component. However,
this ambiguity is not expected to exceed several 10\% greatly 
because of the confirmed tracking accuracy of IRTF SpeX. Hence, our
slit spectra can provide a {\it rough} flux estimate for the compact
emission.

\subsection{Millimeter Interferometric Observations}

Millimeter interferometric observations of HCN (J = 1--0) ($\lambda_{\rm
rest}$ = 3.3848 mm or $\nu_{\rm rest}$ = 88.632 GHz in the rest-frame)
and HCO$^{+}$ (J = 1--0) ($\lambda_{\rm rest}$ = 3.3637 mm or $\nu_{\rm
rest}$ = 89.188 GHz) lines were performed with the Nobeyama Millimeter
Array (NMA) at Nobeyama Radio Observatory (NRO). The NMA consists of
six 10-m antennas. NMA observations were performed in the D (with
longest baseline of 82 m), C (163 m), and AB (351 m) configurations.
The observing log is summarized in Table 2.

The backend was the Ultra-Wide-Band Correlator, UWBC \citep{oku00},
which was configured to cover 1024 MHz with 128 channels at 8-MHz
resolution. The central frequency was set to be 87.99 GHz, to cover
the HCN and HCO$^{+}$ lines simultaneously, redshifted to $\sim$87.75
GHz and $\sim$88.30 GHz, respectively. The bandwidth of 1024 MHz
corresponds to $\sim$3500 km s$^{-1}$ at $\nu$ $\sim$ 88 GHz. The
field of view is $\sim$77 arcsec at this frequency. Since the Hanning
window function was applied to reduce side lobes in the spectra, the
actual resolution was broadened to 16 MHz or 55 km s$^{-1}$ at $\nu$
$\sim$ 88 GHz. To calibrate the passband across the 128 channels, the
bright quasar 0420$-$014 was observed at the beginning of each
observing day. The quasar 1150+497 was used to calibrate temporal
variations in the visibility amplitude and phase.

Standard data reduction was performed using the package UVPROC-II
developed at NRO \citep{tsu97}. For all the data, antenna baselines,
band-pass properties, and time variations of the visibility amplitude
and phase were corrected. Data taken during some parts of
observing time showed large phase scatters, owing to bad radio seeing.
These useless datasets were removed from our analysis. After clipping
a small fraction of unusually high amplitude data, the data were
Fourier-transformed using a natural {\it uv} weighting. The
flux-calibration of Arp 299 was made by observing the quasar 1749+096,
whose flux level relative to Uranus and Neptune had been measured. A
conventional CLEAN method was applied to deconvolve the synthesized
beam pattern. Since Arp 299 consists of distinct merging components
and thus shows a spatially extended structure, the primary beam pattern
of the NMA antenna was corrected. The total net on-source integration
time of Arp 299 was $\sim$25 h. The synthesized beam size was
4$\farcs$2 $\times$ 3$\farcs$8 (position angle is 26$^{\circ}$ west of
north), and the absolute positional uncertainty of the NMA map is well
below 1 arcsec.

\section{Results}

\subsection{Infrared Spectra of Arp 299 A, B1, and C}

Figures 1 and 2 present, respectively, the infrared $L$- and $K$-band
spectra of Arp 299 A, B1, and C. Although Arp 299 B1 and B2 were
simultaneously observed, and signals from both the nuclei were
recognizable, B2 turned out to be much fainter than B1. In the
literature, the nuclear flux of B1 has been reported to be only a
factor of 5 and 1.6 brighter than B2 in the $L$- and $K$-bands,
respectively, when measured with a 1--2-arcsec aperture
\citep{zho93,soi01}. However, the contrast of the nuclear $L$- and
$K$-band flux of B1, relative to B2, is nearly an order of magnitude
in our spectra. Reliable spectral extraction of the much fainter B2
component was therefore very difficult. Although we do not show
infrared $L$- and $K$-band spectra of B2 in this paper, it is less
likely that a luminous buried AGN is present in B2 because the
small B2 to B1 flux ratio at 10 $\mu$m \citep{soi01} suggests that
dust emission heated by a putative buried AGN in B2 is weak, and 
the infrared $K$-band spectrum of B2 is typical of a starburst galaxy,
with no explicit AGN signatures \citep{shi96,alo00}.

\subsubsection{$L$-band spectra}

The 3.3 $\mu$m PAH emission feature is detected clearly at the
infrared continuum peaks of the A, B1, and C components (Figure 1), 
suggesting the presence of a large amount of starburst activity there. 
To estimate the 3.3 $\mu$m PAH strength, we adopt a template spectral
shape for the Galactic star-forming regions and nearby starburst
galaxies (type-1 sources; Tokunaga et al. 1991), as we have done
previously for other LIRGs \citep{idm06,ima06}. The rest-frame 3.3
$\mu$m PAH emission equivalent widths we measured are summarized
in column 2 of Table 3. Table 4 (column 2) tabulates the observed flux
of the 3.3 $\mu$m PAH emission from the infrared continuum peaks. As
noted in $\S$2.1, unlike the equivalent width, the PAH {\it flux} can
be subject to the effects of possible slit loss, even for spatially
compact components, and so our flux estimates are only approximate.

The $L$-band spectrum of B1 displays an absorption feature at the longer
wavelength side of the 3.3 $\mu$m PAH emission feature. We attribute
this feature to 3.4 $\mu$m bare carbonaceous dust absorption, as seen
in highly reddened Galactic stars (e.g., Pendleton et al. 1994;
Imanishi et al. 1996; Rawlings et al. 2003) and obscured AGNs (e.g.,
Imanishi 2000a,b; Imanishi et al. 2001; Risaliti et al. 2003; Mason et
al. 2004; Imanishi et al. 2006a). We estimate the optical depth of the
3.4 $\mu$m absorption feature to be $\tau_{3.4}$ $\sim$ 0.06, which
corresponds to A$_{\rm V}$ $ = $ 8--15 mag, if the Galactic
$\tau_{3.4}$/A$_{\rm V}$ ratio is assumed (= 0.004--0.007; Pendleton et
al. 1994).

A broad absorption feature at $\sim$3.1 $\mu$m is seen in the $L$-band
spectrum of C. We ascribe this feature to 3.1 $\mu$m H$_{2}$O
absorption, as observed in Galactic stars inside or behind molecular
clouds (e.g., Smith et al. 1989, 1993; Murakawa et al. 2000) and LIRGs
\citep{im03,idm06,ima06}. The absorption optical depth is estimated to
be $\tau_{3.1}$ $\sim$ 1.4.

\subsubsection{$K$-band spectra}

The $K$-band spectra in Figure 2 show emission lines of atomic and
molecular (H$_{2}$) hydrogen in the Arp 299 A, B1, and C components.
Table 5 summarizes the rest-frame equivalent widths of the H$_{2}$
S(1) ($\lambda_{\rm rest}$ = 2.122 $\mu$m) and Br$\gamma$
($\lambda_{\rm rest}$ = 2.166 $\mu$m) emission lines, and their flux
ratios, both of which are relatively insensitive to the possible slit
loss.

The Arp 299 A spectrum shows a clear gap in the continuum at
$\lambda_{\rm obs} \sim$ 2.3 $\mu$m in the observed frame, which we
attribute to CO absorption features at $\lambda_{\rm rest}$ =
2.31--2.4 $\mu$m produced by stars \citep{iva00,ia04,imw04}. To
estimate the CO absorption strengths in A, B1, and C, we adopt the
spectroscopic CO index (CO$_{\rm spec}$) defined by \citet{doy94} and
follow the procedures applied to other LIRGs \citep{ima04,ima06}.
Power-law continuum levels (F$_{\rm \lambda}$ = $\alpha \times
\lambda^{\beta}$) are determined using data points at $\lambda_{\rm
obs}$ = 2.07--2.31 $\mu$m ($\lambda_{\rm rest}$ = 2.05--2.29 $\mu$m),
excluding obvious emission lines. The adopted continuum levels are
shown as dashed lines in Figure 2. Data at $\lambda_{\rm obs}$ =
2.33--2.42 $\mu$m ($\lambda_{\rm rest}$ = 2.31--2.40 $\mu$m) are used
to derive the CO$_{\rm spec}$ values. We obtain values of CO$_{\rm
spec}$ $\sim$0.25 (A), $<$0.03 (B1), and $\sim$0.1 (C).

\subsection{Millimeter Interferometric Data}

\citet{aal97} and \citet{cas99} have previously published millimeter
interferometric maps of Arp 299 in the HCN (J = 1--0) line and nearby
continuum; however, the HCO$^{+}$ (J = 1--0) line was not covered
because of a relatively narrow frequency coverage. Ours are the first
interferometric maps of Arp 299 in which the HCN and HCO$^{+}$ lines
were simultaneously observed.

Spectra at the peak positions of A, B, C, and C' showed that the flux
levels between the HCN (J = 1--0) and HCO$^{+}$ (J = 1--0) emission
lines are substantially above the zero level, suggesting that
continuum emission is present. We combined data points that were
unaffected by the HCN and HCO$^{+}$ lines and made interferometric
maps of the continuum emission at individual merging components.
Figure 3 presents the contours of the continuum emission. The continuum
emission is clearly detected in A, B (B1+B2), C, and C' and is
slightly spatially extended, compared to the beam pattern, suggesting
the presence of extended starbursts. A marginal indication of
continuum emission exists, possibly associated with D in our map as well as
in the map of \citet{cas99}, but its presence needs further
confirmation. Although our millimeter interferometric data cannot
resolve the B nucleus clearly into B1 and B2 with $\sim$2.5-arcsec
separation \citep{wyn91}, most of the continuum emission appears to
come from B1 (Figure 3). The estimated $\sim$88 GHz continuum flux
levels for each component are summarized in Table 6.

After subtracting the continuum, we investigate the spatial
distribution of the HCN and HCO$^{+}$ emission lines. Figures 4 and 5
show the integrated intensity map and channel map of the HCO$^{+}$ and
HCN emission lines, respectively. HCO$^{+}$ emission is clearly
detected in the A and B nuclei, and the C-C' region. 
Compared to the beam size, slight spatial extension is found
in component A, but the HCO$^{+}$ emission shows no major
spatial extent in B. In the C region, the HCO$^{+}$ emission is
clearly spatially extended, and shows a peak between C and C',
which we name C2. The HCN emission peaks at the A nucleus, where it is
spatially unresolved. No clear HCN emission is found at B, C, or C' in
our integrated intensity map, but some HCN emission may be present
between C and C'. 

Figure 6 shows continuum-subtracted spectra at the continuum peak
positions of A, B, and C, and HCO$^{+}$ peak C2. 
For B, C, and C2, flux excesses are seen at the expected HCN 
positions in the spectra, which we attribute to HCN emission. 
Figure 7 shows the  
Gaussian fits of the detected HCN and HCO$^{+}$ emission lines at
these peaks. The optical LSR velocity (v$_{\rm opt}$ 
$\equiv$ ($\frac{\lambda}{\lambda_0}$ $-$ 1) $\times$ c; c is the
light speed) and FWHM of the emission lines are summarized in Table 7.
The integrated fluxes of HCN and HCO$^{+}$ at these peaks are derived
from the Gaussian fits of the line profiles and are summarized in
Table 8. For the spatially unresolved HCN emission from A and the
HCO$^{+}$ emission from B, the integrated fluxes are also estimated
from the peak values of the contours of the integrated intensity maps
in Figure 4. The estimated values are consistent with each other within
20\%.

The HCN/HCO$^{+}$ ratios in brightness temperature ($\propto$
$\lambda^{2}$ $\times$ flux density) at the A, B, C, and C2 peaks are
also summarized in Table 8. Although the {\it absolute} flux levels of
the NMA data are uncertain to within $\sim$20\%, this uncertainty does
not propagate to the HCN/HCO$^{+}$ {\it ratios} in brightness temperature because both lines were taken simultaneously with the
same receiver and same correlator unit, under the same weather
conditions, and were calibrated with the same procedure.

To take into account the flux of the extended components, we estimate
the fluxes of HCO$^{+}$ and HCN emission integrated over the area A,
B, and C+C', based on the integrated intensity map in Figure 4. Table
9 tabulates the estimated values. For the HCN emission from A and
HCO$^{+}$ emission from B, the values from the integrated areas are
similar to those for the peak positions, supporting the previous
argument that the emission is spatially unresolved. However, for the
HCO$^{+}$ emission from A and C+C', the values are larger than those
at the peaks, which is consistent with the presence of major
spatially extended components, as was implied from the maps in Figure
4.

\section{Discussion}

\subsection{Comparison to Previously Obtained Data}

\subsubsection{Infrared $L$-band spectra}

The 3.3 $\mu$m PAH emission from Arp 299 has been detected previously
by several groups \citep{miz94,den90,sat99,imd00}. The equivalent
widths of the 3.3 $\mu$m PAH emission (EW$_{\rm 3.3PAH}$) measured
from these older data are tabulated in Table 3 and are compared to
our data.

For the Arp 299 A nucleus, the EW$_{\rm 3.3PAH}$ values are similar
for all the recently obtained data (this work; Satyapal et al. 1999;
Imanishi \& Dudley 2000). Older data presented by \citet{den90} and
\citet{miz94} show smaller values, which could be attributed to larger
scatter and narrower wavelength coverage for their spectra compared to
the more recent observations.

For Arp 299 C, the EW$_{\rm 3.3PAH}$ values reported here and by
\citet{miz94} are a factor of 3 lower than that derived by
\citet{sat99}. This difference can naturally be explained by the
strong 3.1 $\mu$m ice-covered dust absorption feature detected in our
$L$-band spectrum. Because of the flux depression caused by this
absorption feature, the actual continuum level is better derived from
data at $\lambda$ $>$ 3.4 $\mu$m. The use of a data point at $\lambda$
= 3.159 $\mu$m \citep{sat99} can lead to a major overestimate of
the 3.3 $\mu$m PAH emission strength.

For Arp 299 B1, our EW$_{\rm 3.3PAH}$ value is substantially lower
than those derived by \citet{sat99} and by \citet{miz94}. Unlike Arp
299 C, no strong 3.1 $\mu$m absorption feature is detected in B1. We
believe that this EW$_{\rm 3.3PAH}$ difference is likely to be caused
by an increased continuum flux in our data compared to the older data,
for the following reasons. First, the $L$-band continuum flux level is
more than 1 mag brighter than the value from the literature
\citep{zho93}. Although our slit spectra have some flux ambiguity
($\S$2.1), this difference in B1 is difficult to account for solely by
this ambiguity because the $L$-band continuum flux levels for A and C
are similar to those from the same study \citep{zho93}. Next, the B1
and B2 nuclei were simultaneously observed in our spectra, and yet the
$L$-band continuum level in B1 relative to B2 is more than 1 mag
brighter than expected \citep{zho93}. Unless the B2 flux has decreased
substantially, a flux increase in B1 is strongly suggested.

In Table 4, the 3.3 $\mu$m PAH emission fluxes measured in previously
obtained data are shown and compared with our data. We mainly use
the highest quality recent data taken by \citet{sat99} for comparison
with our data. In the A nucleus, both datasets provide similar fluxes.
For the B nucleus, the bulk of the 3.3 $\mu$m PAH emission comes from
B1 \citep{sat99}, and our estimate (B1) and that by Satyapal et al.
(B1+B2) are similar to each other. In the C region, the PAH emission
is extremely spatially extended \citep{sat99}, and the lower flux
in our dataset can be explained by our smaller aperture, in addition
to the above-mentioned PAH flux overestimate from their data. Overall,
the comparison of PAH flux measurements suggests that our absolute
flux calibration is reasonable, and thus the $L$-band continuum flux
increase in B1 as implied above should be real.

\subsubsection{Infrared $K$-band spectra}

Some parts of the $K$-band spectra of the individual merging
components of Arp 299 have been previously obtained
\citep{nak89,rid94,shi94,sug99,alo00}. For the A and C components, the
CO$_{\rm spec}$ values, as well as the emission properties of the
H$_{2}$ S(1) and Br$\gamma$ (Table 5), are similar for our results and
older spectra \citep{nak89,rid94,shi94,sug99}. For the B1 nucleus,
however, the rest-frame equivalent widths of both H$_{2}$ S(1) and
Br$\gamma$ emission lines in our data (Table 5) are a factor of
$\sim$3 smaller than those estimated from older data \citep{alo00}.
Our estimate of CO$_{\rm spec}$ ($<$ 0.03) is also substantially lower
than previous estimates (0.09--0.14; Ridgway et al. 1994;
Alonso-Herrero et al. 2000). Finally, while the $K$-band continuum
flux level for C is similar for our data and older spectra
\citep{alo00}, that for B1 is brighter in our data. All of these
results for B1 support the argument that the flux of the featureless
$K$-band continuum has increased markedly and dilutes the emission and
absorption features in the $K$-band spectrum of B1.

\subsubsection{Millimeter interferometric data}

For the continuum and HCN emission, our results can be compared to
previously obtained interferometric maps \citep{aal97,cas99} and to
single-dish data \citep{sol92}.

The continuum flux level was measured by \citet{cas99} at $\nu$ = 87.7
GHz to be 23.5 (A), 6 (B), 3 (C), and 4 mJy (C'). Our measurements at
$\nu \sim$ 88 GHz (Table 6) are similar to these for all components.

The HCN emission peaks at the A nucleus in our map, corroborating
previous interferometric maps \citep{aal97,cas99}. The
integrated HCN flux at the peak of the A nucleus (Table 8) and an
integrated area surrounding it (Table 9) are similar within $\sim$30\%
to the estimate by Aalto et al. (6.4 and 7 Jy km s$^{-1}$,
respectively). In the map of \citet{cas99}, clear HCN emission is
detected in B, C, and C', while it is not in our map. In the HCN
integrated intensity map of \citet{cas99}, the noise level is lower
than ours, and the highest contour at the B1 nucleus is 1.2 Jy km
s$^{-1}$ beam$^{-1}$, which is lower than the 3$\sigma$ level of our
map. For the HCN emission from C and C', the overall spatial
distribution is similar for our maps and those of \citet{cas99}. In
summary, the distribution of HCN emission from the whole Arp 299
merging system is consistent for all the available interferometric
maps.

The linewidths are 310 km s$^{-1}$ and 280 km s$^{-1}$ in FWHM for HCN
and HCO$^{+}$ emission, respectively, in the A nucleus. These values
agree to within 20\% with the $^{12}$CO linewidth (350 km s$^{-1}$;
Aalto et al. 1997). In many LIRGs, the line profiles of HCN and
$^{12}$CO have been found to be similar \citep{gao04,ink06}, and our
data show this to be the case for the Arp 299 A nucleus. For the Arp
299 B and C components, the FWHM value is better determined from
HCO$^{+}$ than HCN in our data because of the much higher S/N ratios
in the former. The HCO$^{+}$ linewidths in FWHM are 230 km s$^{-1}$
(B) and 130 km s$^{-1}$ (C). For comparison, the $^{12}$CO linewidths
in FWHM are 260 km s$^{-1}$ (B) and 60--80 km s$^{-1}$ (C)
\citep{aal97}.

Finally, our new HCO$^{+}$ interferometric map, the first to be made
for Arp 299, is compared to previous $^{12}$CO maps
\citep{sar91,aal97,cas99}. At the A nucleus, an extended tail toward
the southeast direction is recognizable in both HCO$^{+}$ and
$^{12}$CO maps. For the C' component, an extended HCO$^{+}$ emission
component is detected on the eastern side of the C' peak, as was found in
$^{12}$CO and $^{13}$CO maps \citep{sar91,aal97,cas99}. However, while
the $^{12}$CO emission peaks at C and C', the HCO$^{+}$ emission, as
well as the HCN emission \citep{cas99}, has a peak between the C and
C' components, suggesting that high-density (n$_{\rm H2}$ $>$ 10$^{4}$
cm$^{-2}$) molecular gas, as probed by HCN and HCO$^{+}$, is more
concentrated in the in-between region than the diffuse (n$_{\rm H2}$
$\sim$ 10$^{2-3}$ cm$^{-2}$) gas probed by $^{12}$CO.

\subsection{The Location of the Buried AGN--Implications from Infrared Spectra}

The infrared luminosity of Arp 299 is 2 $\times$ 10$^{45}$ ergs
s$^{-1}$ (Table 1). The measured 3.3 $\mu$m PAH emission flux from the
whole Arp 299 merging system (A+B+C) is $\sim$1.8 $\times$ 10$^{-12}$
ergs s$^{-1}$ cm$^{-2}$ \citep{sat99}, which corresponds to a 3.3
$\mu$m PAH luminosity of 3.5 $\times$ 10$^{41}$ ergs s$^{-1}$, with
the caveat that the actual value is slightly smaller 
because of the overestimate in C ($\S$4.1.1). If we adopt a 3.3 $\mu$m
PAH to infrared luminosity ratio of $\sim$10$^{-3}$, typical for starbursts
\citep{mou90,ima02}, then the luminosity of detected starbursts
becomes 3.5 $\times$ 10$^{44}$ ergs s$^{-1}$, which is a factor of 5
below the measured infrared luminosity. Although the extinction
correction of the 3.3 $\mu$m PAH emission can increase the intrinsic
starburst luminosity, the correction factor is only a factor of 2 to 3
for dust extinction with A$_{\rm V}$ $<$ 20 mag measured in Arp 299
\citep{nak89,sat99,alo00}, if the Galactic dust extinction curve is
assumed \citep{rie85}. Taken at face value, it is quite possible that
AGN activity can play an important role energetically. In this
section, we investigate in detail the nuclei containing buried AGN.

\subsubsection{Arp 299 B1}

The infrared $L$- and $K$-band spectra of the B1 nucleus favor the
presence of a buried AGN there, based on the following three main
arguments. First, the equivalent width of the 3.3 $\mu$m PAH emission in
the $L$-band spectrum is only EW$_{\rm 3.3PAH}$ = 7 nm, more than an
order of magnitude below typical values found in starburst galaxies
(EW$_{\rm 3.3PAH}$ $\sim$ 100 nm: Moorwood 1986; Imanishi \& Dudley
2000). Second, the CO index in the $K$-band spectrum is CO$_{\rm spec}$
$<$ 0.03, much lower than typical values found in starburst galaxies
(CO$_{\rm spec}$ $>$ 0.15; Goldader et al. 1997a,b; Ivanov et al.
2000). A strong contribution from a featureless continuum, most likely
originating from hot dust heated by an AGN, dilutes both the 3.3 $\mu$m
PAH emission and 2.3 $\mu$m CO absorption features. Third, an increase in
the $L$- and $K$-band continua is clearly recognizable, compared to
old data taken about 5--10 years ago. Since no major flux
variation is expected in a starburst over this time period, an AGN is
the natural explanation. We conclude that the B1 nucleus contains a
buried AGN, as was previously argued based on observations of 2--10 keV
X-ray emission \citep{zez03,bal04} and of the infrared 5--16 $\mu$m band
\citep{gal04}. Assuming a Galactic extinction curve of 
$\tau_{3.4}$/A$_{\rm V}$ = 0.004--0.007 \citep{pen94,ima96} and
A$_{\rm L}$ $\sim$ 0.06 $\times$ A$_{\rm V}$ \citep{rie85}, we can
roughly estimate the dereddened luminosity of AGN-heated dust in the
B1 nucleus from the observed $L$-band spectrum (optical depth of dust
absorption features and observed continuum flux level at 3 to 4 $\mu$m)
\citep{idm01,im03,idm06}, which is 3--5 $\times$ 10$^{43}$ ergs
s$^{-1}$.

\subsubsection{Arp 299 C}

Our infrared $L$-band spectrum of C displays a strong 3.1 $\mu$m ice
absorption feature with $\tau_{3.1}$ $\sim$ 1.4. For a buried AGN
in which the energy source is more centrally concentrated than dust, a
foreground-screen dust-absorption geometry is applicable, while a
mixed dust/source geometry is at work for a normal starburst in which
stellar energy sources and dust are spatially well mixed
\citep{pux91,mcl93,for01}. In a normal starburst with mixed
dust/source geometry, whose radiation density, and therefore the
fraction of dust covered with an ice mantle, is similar to the
well studied starburst galaxy M82, the $\tau_{3.1}$ value is $<$0.3
for any amount of dust extinction of the starburst \citep{im03}. Even
if all dust is covered with an ice mantle ($f$ = 1 in Imanishi \&
Maloney 2003), the mixed dust/source geometry can only provide a
maximum of $\tau_{3.1}$ $\sim$ 0.7, assuming the dust properties are
similar to the Galactic interstellar medium. This is because the
observed flux is dominated by foreground, weakly obscured, less
attenuated emission, with a small contribution from obscured, highly
attenuated emission at the far side, in this mixed dust/source
geometry. The large $\tau_{3.1}$ value in the Arp 299 C component
($\sim$1.4) would require a foreground-screen dust geometry. A
centrally concentrated energy source, such as a buried AGN, may be
present, in addition to a normal, PAH-emitting starburst (Figure 1).
Alternatively, if a large amount of foreground-screen dust exists
exterior to the normal starburst core (e.g., in an edge-on host
galaxy), then the large $\tau_{3.1}$ can be explained. However,
the EW$_{\rm 3.3PAH}$ value is $\sim$40 nm in C. In this geometry, the
EW$_{\rm 3.3PAH}$ value should have the value usually found for normal
starbursts ($\sim$100 nm) because both PAH and continuum emission are
attenuated in a similar manner. Dilution of the PAH emission by
PAH-free continuum from an energy source other than a normal starburst
is indicated from the $L$-band spectrum of C. In the C component, the
observed CO$_{\rm spec}$ value ($\sim$0.1) is also smaller than
expected for a normal starburst ($>$0.15). Dilution by the featureless
$K$-band continuum, possibly from AGN-heated hot dust or stars younger
than $\sim$10$^{6}$ yr \citep{lei99}, is required to explain this result.

Unlike the A and B nuclei, Arp 299 C+C$'$ is thought to be a
non-galaxy nucleus \citep{dud93,cha02}. Under the assumption of an
Eddington mass accretion rate, a more massive black hole is necessary
to produce a larger AGN luminosity, and supermassive black holes are
generally present at galaxy nuclei. Whether the C component possesses
a sufficiently large supermassive black hole for AGN activity is less
clear than for the other Arp 299 nuclei. \citet{alo00} noticed a
weaker CO absorption feature in the $K$-band spectrum of C, but
attributed its weakness to the youth of the starbursts. If a very
active super star cluster with an infrared luminosity of L$_{\rm IR}$
$\sim$ 10$^{9}$L$_{\odot}$ \citep{gor01} is formed in the innermost
regions of dense molecular gas and the radiation density there is so
high as to destroy PAH molecules \citep{sel81}, then a large
$\tau_{3.1}$ value and PAH-free featureless $L$-band continuum
\citep{alo04} may be produced. For the nuclei of ultraluminous infrared
galaxies with L$_{\rm IR}$ $>$ 10$^{12}$L$_{\odot}$, energy sources that
are more centrally concentrated than dust would require
both large luminosity and high surface brightnesses, for which AGNs
are the natural explanation \citep{soi00,soi03}. However, the
infrared luminosity of Arp 299 C is at the level of L$_{\rm IR}$
$\sim$10$^{10}$L$_{\odot}$ \citep{cha02}. In Arp 299 C, the scenario
of multiple young super star clusters embedded in dust and molecular
gas may hold in term of absolute luminosity.

\subsubsection{Arp 299 A}

Among the merging components of Arp 299, the A nucleus (IC 694) has
long been suggested to be the most probable AGN-harboring nucleus,
based on near-infrared color \citep{nak89}, compact high
surface-brightness radio emission \citep{lon92,lon93}, and starburst
modeling \citep{geh83,shi96}. However, our infrared $L$- and $K$-band
spectra reveal no explicit AGN evidence in this nucleus: its EW$_{\rm
3.3PAH}$ and CO$_{\rm spec}$ values are typical of starburst galaxies
\citep{moo86,gol97a,gol97b,iva00}. Recent high-quality infrared data
\citep{imd00,cha02,gal04} have also failed to provide clear AGN
signatures from A. Although \citet{bal04} inferred the presence of an
AGN in the A nucleus, based on the presence of 6.7 keV ionized iron
emission, their argument is weaker than that based on the 6.4-keV iron
K$\alpha$ line for the B1 nucleus. If a luminous buried AGN is present
in the Arp 299 A nucleus, the AGN must be more elusive than in B1.

\subsection{Interpretation of the HCN/HCO$^{+}$ Ratios}

\citet{koh01} and \citet{koh05} proposed that AGN-dominated galaxy
nuclei (XDRs) tend to show higher HCN (J = 1--0) to HCO$^{+}$ (J = 1--0)
and HCN (J = 1--0) to CO (J = 1--0) brightness-temperature ratios than
starburst-dominated galaxy nuclei (PDRs), and thus that these ratios
can be used to distinguish the two types of galaxies. In particular,
both the HCN and HCO$^{+}$ molecules have similarly large dipole
moments ($\mu$ $\sim$ 3--4 debye; Millar et al. 1997; Botschwina et
al. 1993) and so probe similarly high-density (n$_{\rm H2}$ $>$
10$^{4}$ cm$^{-2}$) molecular gas. Hence, the HCN/HCO$^{+}$ ratios are
very useful for distinguishing between the two galaxy types, being
unaffected by different dense-to-diffuse molecular gas fractions in
different galaxies. Imanishi et al. (2004; 2006b) applied this method to
a few LIRG nuclei, and successfully found high HCN/HCO$^{+}$
brightness-temperature ratios in LIRGs known to possess luminous
buried AGNs based on previously obtained data at other wavelengths. An
HCN {\it abundance} enhancement in XDR chemistry of dense molecular gas
around a strong X-ray illuminating source (Meijerink \& Spaans 2005;
their Fig.10, model 4) can naturally explain the high HCN/HCO$^{+}$
brightness-temperature ratios in luminous AGNs, whether the HCN and
HCO$^{+}$ emission is optically thin or thick \citep{ink06}. 

We use the HCN and HCO$^{+}$ flux measurements obtained with the NMA,
to apply this energy diagnostic method to the individual merging
components of Arp 299. For Arp 299, HCN and HCO$^{+}$ measurements
using a single-dish radio telescope are available \citep{gra06}.
However, our interferometric data have the following advantages over
the single-dish measurements: Although single-dish radio data
measure the HCN and HCO$^{+}$ emission from the combined B+C regions,
our interferometric data can clearly resolve B and C because of
better spatial resolution. In addition, we can
probe the dense molecular gas for regions confined to the nuclei where
putative AGNs are expected to reside, by reducing the contamination from
extended starburst emission. The XDR volume can be large in a {\it pure}
buried AGN with no starburst activity. However, in a
starburst/buried-AGN composite 
galaxy, such as the majority of LIRGs, strong XDR signatures are
expected only in the nuclear region in the close vicinity of a
luminous X-ray-emitting AGN. Spatially extended molecular gas can be
affected by the surrounding starbursts and thus show properties
expected from PDRs. For the purpose of finding XDR signatures in
starburst/AGN composite LIRGs, such as Arp 299, investigating the
HCN/HCO$^{+}$ ratios in regions confined to the nuclei, based on
high-spatial-resolution interferometric data, is superior to using
single-dish data. Furthermore, in the old single-dish data, HCN and
HCO$^{+}$ observations were made at different times, whereas for our NMA
interferometric data, the HCN and HCO$^{+}$ lines were simultaneously
obtained, so that possible uncertainties due to time-variable
observing conditions and intercalibration are minimized, making the
discussions of the HCN/HCO$^{+}$ brightness-temperature {\it ratios}
reliable.

Figure 8 shows the HCN (J = 1--0) to CO (J = 1--0) (abscissa) and HCN
(J = 1--0) to HCO$^{+}$ (J = 1--0) (ordinate) ratios in brightness
temperature of Arp 299 and other galaxies. For the Arp 299 merging
components, the HCN/HCO$^{+}$ brightness-temperature ratios are
directly derived from our NMA data, while the HCN/CO ratios are taken
from the older interferometric data of \citet{cas99}. The Arp 299 A, B,
and C components show low HCN/HCO$^{+}$ brightness-temperature ratios,
as observed in starburst-dominated galaxy nuclei.

Table 10 summarizes the detection or non-detection of AGN signatures
based on various methods.  
Since a luminous buried AGN is almost certainly present in the B1
nucleus, we must consider why no XDR signature is found
in B based on the HCN/HCO$^{+}$ ratio. Several explanations are
possible, but whichever scenario is correct, we have to conclude that
the HCN/HCO$^{+}$ method may miss some types of buried AGN despite
their detection at other wavelengths.

First, the dereddened AGN luminosity at the B1 nucleus derived from
our infrared $L$-band spectrum is L$_{\rm IR}$(AGN) $\sim$ 3--5 $\times$
10$^{43}$ ergs s$^{-1}$ ($\S$4.2.1). The absorption-corrected 2--10 
keV X-ray luminosity is $\sim$5 $\times$ 10$^{42}$ ergs s$^{-1}$
\citep{del02}. If we apply 2--10 keV to the bolometric luminosity
correction of $\sim$10--20 \citep{elv94}, the estimated AGN luminosity
becomes roughly similar, within a factor of 2--3, to the value quoted
above based on our infrared $L$-band spectrum. However, the detected starburst
luminosity in the B1 nucleus is estimated to be L$_{\rm IR}$(SB) $\sim$ 4--5
$\times$ 10$^{43}$ ergs s$^{-1}$ based on the observed 3.3 $\mu$m PAH
emission luminosity (This work; Satyapal et al. 1999) and on the
ratio of L$_{\rm 3.3 PAH}$/L$_{\rm IR}$ $\sim$ 10$^{-3}$ found in
starbursts \citep{mou90,ima02}. If we assume dust
extinction of A$_{\rm V}$ = 6--15 mag for the detected starbursts in B
\citep{geh83,bec86,sat99}, the extinction-corrected starburst
luminosity can increase to L$_{\rm IR}$(SB) $\sim$ 7--10 $\times$
10$^{43}$ ergs s$^{-1}$. Thus, in the B nucleus, the starburst
luminosity is comparable to the AGN luminosity. 
The presence of high-density molecular gas strongly affected by this
luminous starburst component may hide the signatures of XDRs traced by
our HCN/HCO$^{+}$ method. 

Second, for the HCN/HCO$^{+}$ method to work in finding XDR
signatures, the high-density molecular gas, as probed by HCN and
HCO$^{+}$, must be sufficiently close to be affected by the
central AGN's X-ray radiation. In the double nuclei merging LIRG NGC
6240, the HCN and HCO$^{+}$ emission peaks are not at individual nuclei,
but in the internuclear region where shocks are important
\citep{nak05}. In Arp 299, although the HCO$^{+}$ emission peak from B
(B1+B2) appears to coincide with the B1 nucleus in our interferometric
map with 4$\farcs$2 $\times$ 3$\farcs$8 spatial resolution (Figure 4),
it is possible that the actual distribution is slightly displaced from
the exact position of the B1 nucleus to the B2 side, and shocks may
play a major role. It is known that shock regions show enhanced
HCO$^{+}$ abundance \citep{dic80}, which can decrease the
HCN/HCO$^{+}$ brightness-temperature ratio, compared to a pure buried
AGN. In fact, shocks driven by starburst superwinds are found to be
prominent in the whole area of the Arp 299 merging system
\citep{hec99}. These spatially extended shocks at the surface of the
buried AGN may decrease the HCN/HCO$^{+}$ brightness-temperature ratio
measured toward the B1 nucleus.

Third, an increase in AGN luminosity compared to older data is
indicated in our infrared spectra of B1 ($\S$4.2.1). Some time lag
may perhaps occur between the onset of an active AGN phase and the stage
when the X-ray-emitting AGN can chemically affect the surrounding 
high-density molecular gas. If the very active AGN phase has just started,
then the low HCN/HCO$^{+}$ brightness-temperature ratio may be
explained if the surrounding molecular gas has not been subjected to
sufficient chemical effects.

Finally, the HCN/HCO$^{+}$ abundance ratio is highly dependent on the
density and distribution of molecular gas surrounding a central
X-ray-emitting AGN \citep{mei05}. The Arp 299 merging components may
have different molecular gas distributions (e.g., density and geometry)
from other LIRGs for which our HCN/HCO$^{+}$ method succeeded. Further
detailed chemical calculations are needed to interpret more
unambiguously the observed HCN/HCO$^{+}$ brightness-temperature ratios
in galaxies. 

\section{Summary}

We performed infrared $L$- and $K$-band spectroscopy of the Arp 299 A,
B1, and C regions to identify the location of the putative
Compton-thick buried AGN, revealed by recent X-ray data. Emission
and absorption features found in the $L$- and $K$-band spectra were
used to distinguish between an AGN and a starburst. Millimeter
interferometric maps of simultaneously taken HCN (J = 1--0) and
HCO$^{+}$ (J = 1--0) molecular lines were also presented for the Arp 299
A, B (B1+B2), and C regions. Using the interferometric maps, we
determined the spatial distribution of these molecules and derived the
HCN/HCO$^{+}$ ratios in brightness temperature, to search for possible
chemical signatures of the putative X-ray-emitting buried AGN to the
surrounding interstellar medium. We obtained the following main
conclusions.

\begin{enumerate}
\item Our infrared spectra strongly suggested that a luminous buried AGN
      is present in the B1 nucleus, based on a very low equivalent
      width of the 3.3 $\mu$m PAH emission in the $L$-band spectrum, 
      a very weak 2.3 $\mu$m CO absorption feature in the $K$-band
      spectrum, and a major increase in the $L$- and $K$-band
      continuum fluxes compared to between 5 and 10 years ago.  
      Our results supported previous arguments that the B1 nucleus
      hosts an AGN.  
      The dereddened AGN luminosity from our infrared $L$-band spectrum
      was quantitatively comparable to that based on X-ray observations 
      at E $>$ 10 keV.  
\item Our infrared spectra of the non-galaxy-nucleus Arp 299 C showed
      a large optical depth of 3.1 $\mu$m ice absorption 
      ($\tau_{3.1}$ $\sim$ 1.4) in the $L$ band, and a weak 2.3 
      $\mu$m CO absorption feature (CO$_{\rm spec}$ $\sim$ 0.1) in the
  	$K$ band. 
      Although a buried AGN is a plausible explanation, multiple 
      young super star clusters are also a possibility, given the small 
      infrared luminosity of Arp 299 C (L$_{\rm IR}$ $\sim$
      10$^{10}$L$_{\odot}$). 
\item In our infrared $L$- and $K$-band spectra of the A nucleus, we
      found no explicit evidence for an AGN.
\item Our millimeter interferometric maps confirmed that 
      continuum emission is detected in the A, B, C, and C' regions and
      that HCN (J = 1--0) emission has a strong peak at the A nucleus. 
      The flux and spatial distribution of the continuum and HCN emission 
      were similar to those from previously obtained interferometric maps.  
\item We presented the first interferometric map of Arp 299 in
      HCO$^{+}$ (J = 1--0) line emission. 
      HCO$^{+}$ emission was clearly detected in the Arp 299 A, B, C, and C'
      regions.
      In the C and C' regions, the HCO$^{+}$ emission peaks between 
      these regions.
      This spatial distribution is similar to that of HCN, but is
      unlike the distribution of
      $^{12}$CO (J = 1--0), which shows peaks in the C and C' components. 
      We conclude that high-density molecular gas, as probed by the HCN
      and HCO$^{+}$ lines, has been swept to the in-between regions, while
      the more diffuse
      gas traced by $^{12}$CO has not.
\item We measured HCN/HCO$^{+}$ ratios in brightness temperature 
      for the A, B (B1+B2), and C regions, and compared the ratios with 
      those of AGN- and starburst-dominated galaxies. 
      The ratios in the three Arp 299 regions, even in the AGN-hosting
      B1 nucleus, were as low as seen in starburst-dominated galaxy 
      nuclei, possibly because of the effects of coexisting starbursts
      and/or shocks.   
\end{enumerate}


We are grateful to the IRTF and NRO staff for their support during our 
observing runs.
We thank S. Ishizuki for valuable discussions on the basics of
interferometric observations of molecular gas. 
M.I. is supported by Grants-in-Aid for Scientific Research (16740117).
M. I. is a visiting astronomer at the Infrared
Telescope Facility, which is operated by the University of Hawaii under
Cooperative Agreement no. NCC 5-538 with the National Aeronautics and
Space Administration, Science Mission Directorate, Planetary Astronomy
Program.
NRO is a branch of the National Astronomical Observatory, National
Institutes of Natural Sciences, Japan.
Some parts of the data analysis were made using a computer system operated
by the Astronomical Data Analysis Center (ADAC) and the Subaru Telescope
of the National Astronomical Observatory.
This publication makes use of data products from the Two Micron All Sky
Survey, which is a joint project of the University of Massachusetts and
the Infrared Processing and Analysis Center/California Institute of
Technology, funded by the National Aeronautics and Space Administration
and the National Science Foundation.

\clearpage

\begin{table}[h]
\caption{Detailed information on Arp 299 \label{tab1}}
\begin{center}
\begin{tabular}{cccccccc}
\hline
\hline
Object & Redshift &  
f$_{\rm 12}$  & f$_{\rm 25}$  & 
f$_{\rm 60}$  & f$_{\rm 100}$  & 
log L$_{\rm IR}$ & log L$_{\rm IR}$/L$_{\odot}$ \\
 &    & (Jy) & (Jy)  & (Jy) 
& (Jy)  & (ergs s$^{-1}$)  \\
(1) & (2) & (3) & (4) & 
(5) & (6) & (7) & (8) \\ \hline
Arp 299 & 0.010 & 3.8 & 23.2 & 103.7 & 107.4 & 45.3 & 11.7 \\
\hline
\end{tabular}
\end{center}
\end{table}

Notes.

(1): Object.

(2): Redshift. 

(3)--(6): f$_{12}$, f$_{25}$, f$_{60}$, and f$_{100}$ are 
{\it IRAS FSC}
fluxes at 12$\mu$m, 25$\mu$m, 60$\mu$m, and 100$\mu$m, respectively.

(7): Decimal logarithm of infrared (8$-$1000 $\mu$m) luminosity
in ergs s$^{-1}$ calculated with
$L_{\rm IR} = 2.1 \times 10^{39} \times$ D(Mpc)$^{2}$
$\times$ (13.48 $\times$ $f_{12}$ + 5.16 $\times$ $f_{25}$ +
$2.58 \times f_{60} + f_{100}$) ergs s$^{-1}$
\citep{sam96}. 

(8): Decimal logarithm of infrared luminosity in units of solar
luminosity. 

\begin{table}[h]
\caption{NMA observing log\label{tab2}}  
\begin{center}
\begin{tabular}{cc}
\hline
\hline
Array & Observing Date \\ 
Configuration & (UT) \\ \hline
D & 2004 Mar 12--14 \\
C & 2003 Dec 9 \\
AB & 2004 Jan 11 \\ \hline
\end{tabular}
\end{center}
\end{table}

\begin{table}[h]
\caption{Rest-frame equivalent width of the 3.3 $\mu$m PAH emission
feature in [nm] \label{tab3}}
\begin{center}
\begin{tabular}{cccccc}
\hline
\hline
Component & This work & S99 & DD90 & M94 & ID00 \\
(1) & (2) & (3) & (4) & (5) & (6) \\ \hline
A  & 150 (0$\farcs$8 $\times$ 3$\farcs$0) & 130 $^{a}$ 
(2$\farcs$7) & 70 (2$\farcs$7) & 60 (5$\farcs$4 $\times$ 7$\farcs$3) & 
150 (1$\farcs$2 $\times$ 10$\farcs$0) \\ 
B1 & 7 (0$\farcs$8 $\times$ 3$\farcs$0) & --- & --- & --- &
--- \\
B (B1+B2) & --- & 14 $^{a}$  (7$\farcs$1) & --- & 40
(5$\farcs$4 $\times$ 7$\farcs$3) & --- \\ 
C  & 40 (0$\farcs$8 $\times$ 3$\farcs$0) & 144 $^{a}$ (7$\farcs$1) & --- &
40 (5$\farcs$4 $\times$ 7$\farcs$3) & --- \\
\hline
\end{tabular}
\end{center}
\end{table}

$^{a}$: \citet{sat99} performed narrowband imaging observations at 3.326
$\mu$m (PAH) and 3.159 $\mu$m (continuum).
The flux level at 3.159 $\mu$m is taken as the representative continuum
level at the wavelength of the 3.3 $\mu$m PAH emission feature.

Notes.

(1): Arp 299 merging component.

(2): This paper. 

(3): \citet{sat99}.

(4): \citet{den90}.

(5): \citet{miz94}.

(6): \citet{imd00}. 

\clearpage

\begin{table}[h]
\caption{3.3 $\mu$m PAH emission flux in 10$^{-16}$ [W
m$^{-2}$] \label{tab4}}  
\begin{center}
\begin{tabular}{cccccc}
\hline
\hline
Component & This work & S99 & DD90 & M94 & ID00 \\
(1) & (2) & (3) & (4) & (5) & (6) \\ \hline
A  & 4.5 (0$\farcs$8 $\times$ 3$\farcs$0) & 3.2 (2$\farcs$7) & 
5.0$\pm$0.9 (2$\farcs$7) & 5.8$\pm$1.6 (5$\farcs$4 $\times$ 7$\farcs$3) & 
6 (1$\farcs$2 $\times$ 10$\farcs$0) \\ 
B1 & 2.0 (0$\farcs$8 $\times$ 3$\farcs$0) & --- & --- & ---
& --- \\
B (B1+B2) & --- & 2.7 (7$\farcs$1) & --- & 7.4$\pm$1.5
(5$\farcs$4 $\times$ 7$\farcs$3) & --- \\ 
C  & 1.5 (0$\farcs$8 $\times$ 3$\farcs$0) & 6.3 (7$\farcs$1) & --- &
4.3$\pm$1.5 (5$\farcs$4 $\times$ 7$\farcs$3) & --- \\
\hline
\end{tabular}
\end{center}
\end{table}

Notes. 

(1): Arp 299 merging component.

(2): This paper. 

(3): \citet{sat99}.

(4): \citet{den90}.

(5): \citet{miz94}.

(6): \citet{imd00}. 

\begin{table}[h]
\caption{Br$\gamma$ and H$_{2}$ S(1) emission lines \label{tab5}}  
\begin{center}
\begin{tabular}{ccccccc}
\hline
\hline
Component & \multicolumn{2}{c}{EW [nm]} &
\multicolumn{2}{c}{EW [nm] (A00)} & S(1)/Br$\gamma$ & 
S(1)/Br$\gamma$ (S99) \\   
 & H$_{2}$ S(1) & Br$\gamma$ & H$_{2}$ S(1) & Br$\gamma$ &  &  \\
(1) & (2) & (3) & (4) & (5) & (6) & (7) \\ \hline
A  & 3      & 3   &  --- & --- & 0.9    & 1.2 \\
B1 & 0.3    & 0.3 &  1       & 0.9     & 1      & ---  \\
C  & $<$1.2 & 4   &  0.5     & 4       & $<$0.3 & 0.12 \\
\hline
\end{tabular}
\end{center}
\end{table}

Notes.

(1): Arp 299 merging component.

(2) and (3): Rest frame equivalent width of H$_{2}$ S(1) 
($\lambda_{\rm rest}$ = 2.122 $\mu$m) and Br$\gamma$ 
($\lambda_{\rm rest}$ = 2.166 $\mu$m) emission lines in [nm],
respectively, derived from our data. 

(4) and (5): Rest frame equivalent width of H$_{2}$ S(1)
and Br$\gamma$ emission lines in [nm], respectively, derived by
\citet{alo00}. 

(6): H$_{2}$ S(1) to Br$\gamma$ flux ratio, derived from our data.

(7): H$_{2}$ S(1) to Br$\gamma$ flux ratio, determined by \citet{sug99}.

\begin{table}[h]
\caption{Continuum flux at the position of each merging component
measured at $\nu$ $\sim$ 88 GHz \label{tab6}}  
\begin{center}
\begin{tabular}{cc}
\hline
\hline
Component & Flux \\ 
          & (mJy) \\ \hline
A & 23.5$\pm$1 \\
B & 6$\pm$1 \\
C & 4.5$\pm$1  \\ 
C$'$ & 4.5$\pm$1 \\
\hline
\end{tabular}
\end{center}
\end{table}

\clearpage

\begin{table}[h]
\caption{Gaussian fitting parameters of HCN and HCO$^{+}$ emission
lines \label{tab7}}   
\begin{center}
\begin{tabular}{ccccc}
\hline
\hline
Component & \multicolumn{2}{c}{LSR velocity} &
\multicolumn{2}{c}{FWHM}  \\   
 & \multicolumn{2}{c}{(km s$^{-1}$)} & 
\multicolumn{2}{c}{(km s$^{-1}$)} \\ 
 & HCN & HCO$^{+}$ & HCN & HCO$^{+}$  \\
(1) & (2) & (3) & (4) & (5)  \\ \hline
A  & 3100 & 3100 & 310 & 280 \\
B  & 3080 & 3150 & 100 & 230 \\ 
C  & 3210 & 3170 & 160 & 130 \\
C2 & 3220 & 3170 & 65  & 125 \\
\hline
\end{tabular}
\end{center}
\end{table}

Notes. 

(1): Arp 299 merging component. 

(2): Optical LSR velocity \{v$_{\rm opt}$ $\equiv$ 
($\frac{\lambda}{\lambda_0}$ $-$ 1) $\times$ c\} of the
HCN emission peak.
For Arp 299, the LSR velocity is 5.9 km s$^{-1}$ larger than the
heliocentric velocity. 
For B, C, and C2, the uncertainty may be large, because HCN emission is
weak. 

(3): Optical LSR velocity of the HCO$^{+}$ emission peak.

(4): Line width of the HCN emission at FWHM.
     For B, C, and C2, the uncertainty may be large.

(5): Line width of the HCO$^{+}$ emission at FWHM.

\begin{table}[h]
\caption{Integrated intensity of HCN and HCO$^{+}$ lines at the peak
positions \label{tab8}}   
\begin{center}
\begin{tabular}{ccccc}
\hline
\hline
Component & HCN & HCO$^{+}$ &
HCN/HCO$^{+}$ & HCN/HCO$^{+}$\\   
 & (Jy km s$^{-1}$) & (Jy km s$^{-1}$) & (this work) & (single dish) \\
(1) & (2) & (3)  & (4)  & (5)  \\ \hline
A  & 8.6  & 10.6 & 0.82 & 0.52 \\
B  & 1.1  &  4.2 & 0.25 & 0.64 (B + C + C2) \\
C  & 1.5  &  3.2 & 0.47 & \\ 
C2 & 0.56 &  2.0 & 0.28 & \\
\hline
\end{tabular}
\end{center}
\end{table}

Notes. 

(1): Arp 299 merging component.

(2): Integrated HCN intensity.
For B, C, and C2, the uncertainty may be large.

(3): Integrated HCO$^{+}$ intensity.

(4): HCN/HCO$^{+}$ ratio in brightness temperature 
($\propto$ $\lambda^{2}$ $\times$ flux density).
The ratio is not affected by possible uncertainties in the absolute flux
calibration.

(5): HCN/HCO$^{+}$ ratio in brightness temperature, measured from single
dish telescope's data \citep{gra06}.

\clearpage

\begin{table}[h]
\caption{Fluxes of HCN and HCO$^{+}$ lines, including spatially extended
components \label{tab9}}   
\begin{center} 
\begin{tabular}{cccc}
\hline
\hline
Component & HCN & HCO$^{+}$ \\   
 & (Jy km s$^{-1}$) & (Jy km s$^{-1}$) \\
(1) & (2) & (3) \\ \hline
A     & 10      & 16 \\
B     & --- &  5 \\
C+C'  & 3       & 12  \\ 
\hline
\end{tabular}
\end{center}
\end{table}

Notes. 

(1): Arp 299 merging component.

(2): Integrated HCN intensity.

(3): Integrated HCO$^{+}$ intensity.

\begin{table}[h]
\caption{AGN signatures at individual major merging components of Arp
299, based on various methods \label{tab10}}   
\begin{center} 
\begin{tabular}{c|ccccc}
\hline
\hline
Method  & A & B & C & Reference \\   
(1)     & (2) & (3) & (4) & (5) \\ \hline
Infrared $K$-band & X & $\bigcirc$ & $\triangle$ & a \\
Infrared $L$-band & X & $\bigcirc$ & $\triangle$ & a \\
HCN/HCO$^{+}$     & X & X          & X           & a \\ \hline 
Optical           & X & X $^{\alpha}$ & X        & b,c,d,e \\
X-ray             & $\triangle$ & $\bigcirc$ & X & f,g\\  
Mid-infrared      & X & $\bigcirc$ & X           & h \\
\hline
\end{tabular}
\end{center}
\end{table}

$\alpha$: \citet{gar06} have recently argued the detection
of AGN signatures through {\it integral field} optical spectroscopy. 

Notes. 

(1): Methods.

(2)-(4): AGN signatures at Arp 299 A, B, and C components.
$\bigcirc$: Strong. $\triangle$: Possible. X: None.

(5): References for AGN signatures. 
(a): This work,
(b): \citet{bus84}, 
(c): \citet{kee85},
(d): \citet{arm89},
(e): \citet{coz98},
(f): \citet{zez03},
(g): \citet{bal04},
(h): \citet{gal04}.

\begin{figure}
\begin{center}
    \FigureFile(80mm,80mm){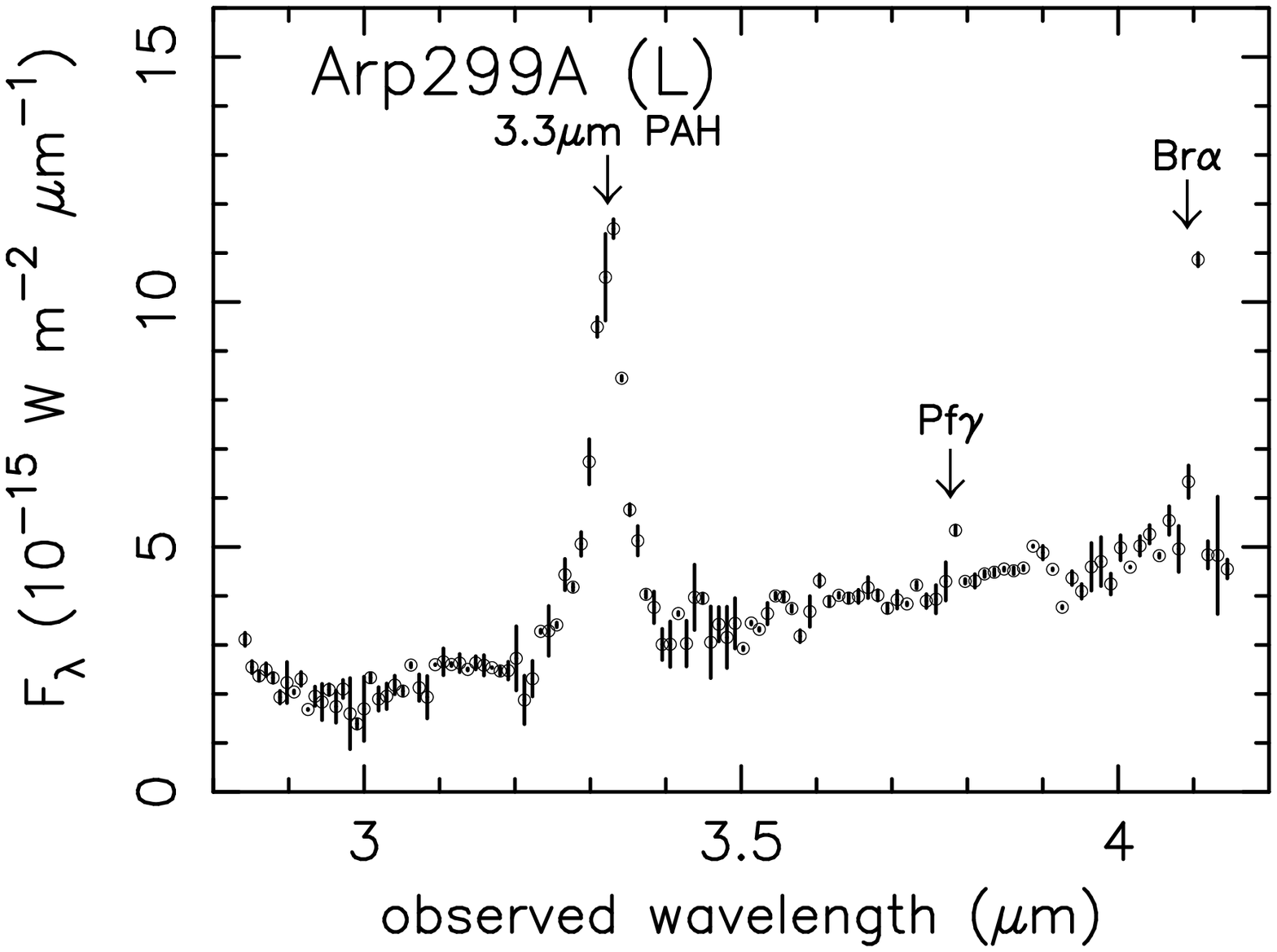}
    \FigureFile(80mm,80mm){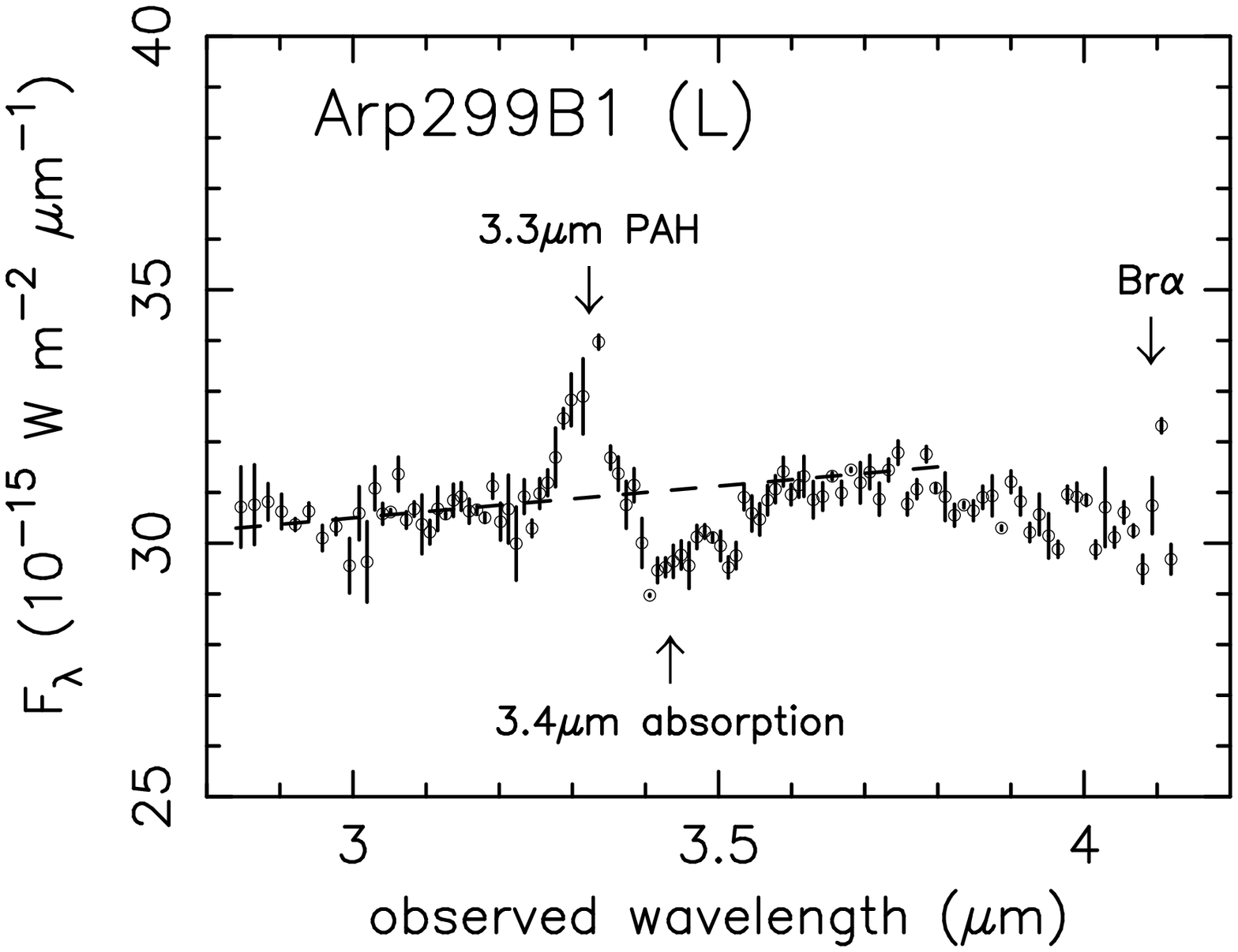} \\
    \FigureFile(80mm,80mm){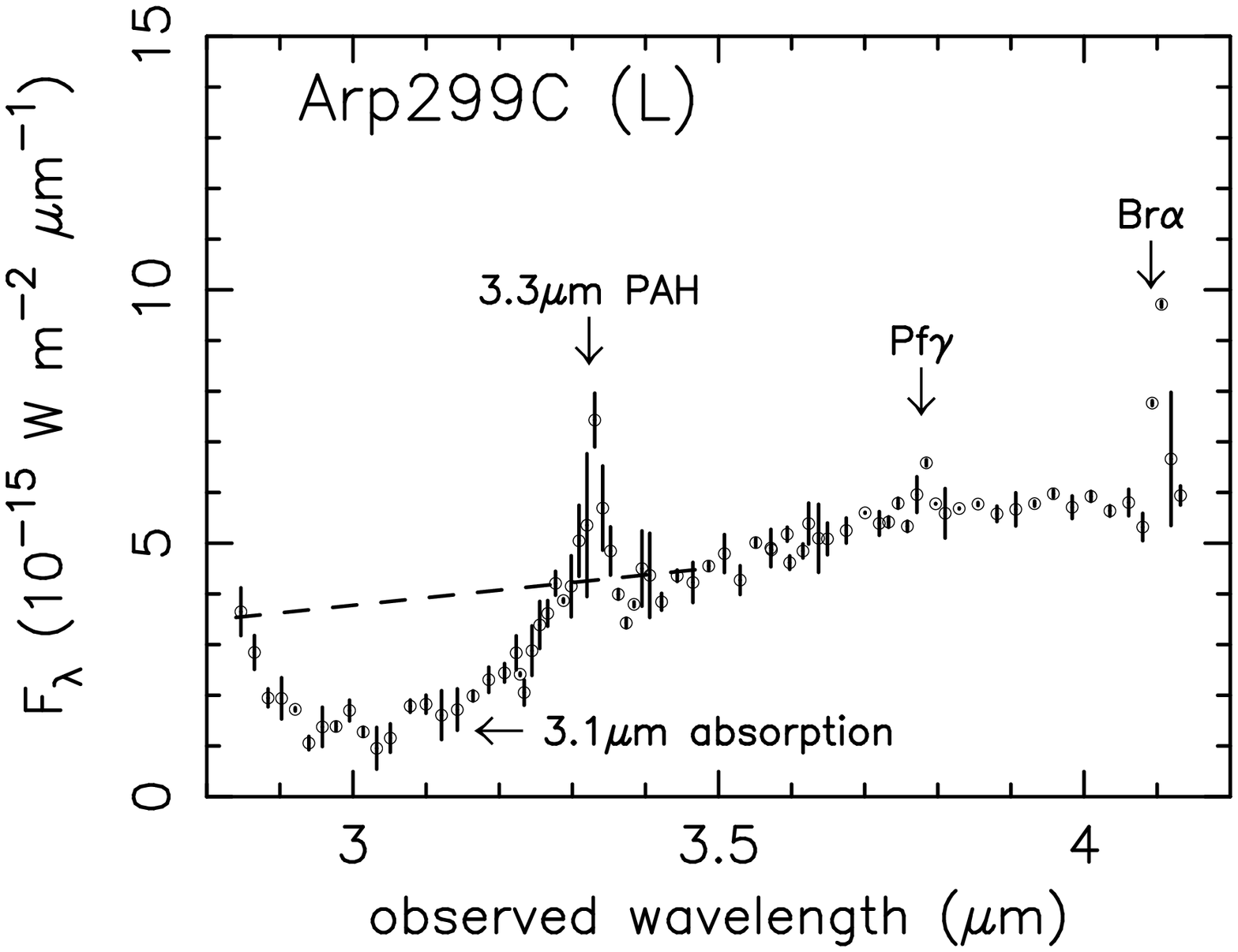}
  \end{center}
\caption{
Infrared $L$-band (2.8--4.1 $\mu$m) spectra of the Arp 299 A, B1, and C 
components.
The abscissa and ordinate are the observed wavelength in $\mu$m and flux
F$_{\lambda}$ in 10$^{-15}$ W m$^{-2}$ $\mu$m$^{-1}$, respectively.  
The dashed lines in the Arp 299 B1 and C spectra are the adopted continuum
levels, which were used to measure the optical depths of 
3.4 $\mu$m bare carbonaceous dust (B1) or 3.1 $\mu$m ice-covered dust
(C) absorption features, respectively.  
In addition to the detected 3.3 $\mu$m PAH emission, detected
emission lines of hydrogen are indicated.   
Br$\alpha$: Br$\alpha$ emission line at $\lambda_{\rm rest}$ = 4.051 $\mu$m.
Pf$\gamma$: Pf$\gamma$ emission line at $\lambda_{\rm rest}$ = 3.738 $\mu$m.
}
\end{figure}

\begin{figure}
 \begin{center}
    \FigureFile(80mm,80mm){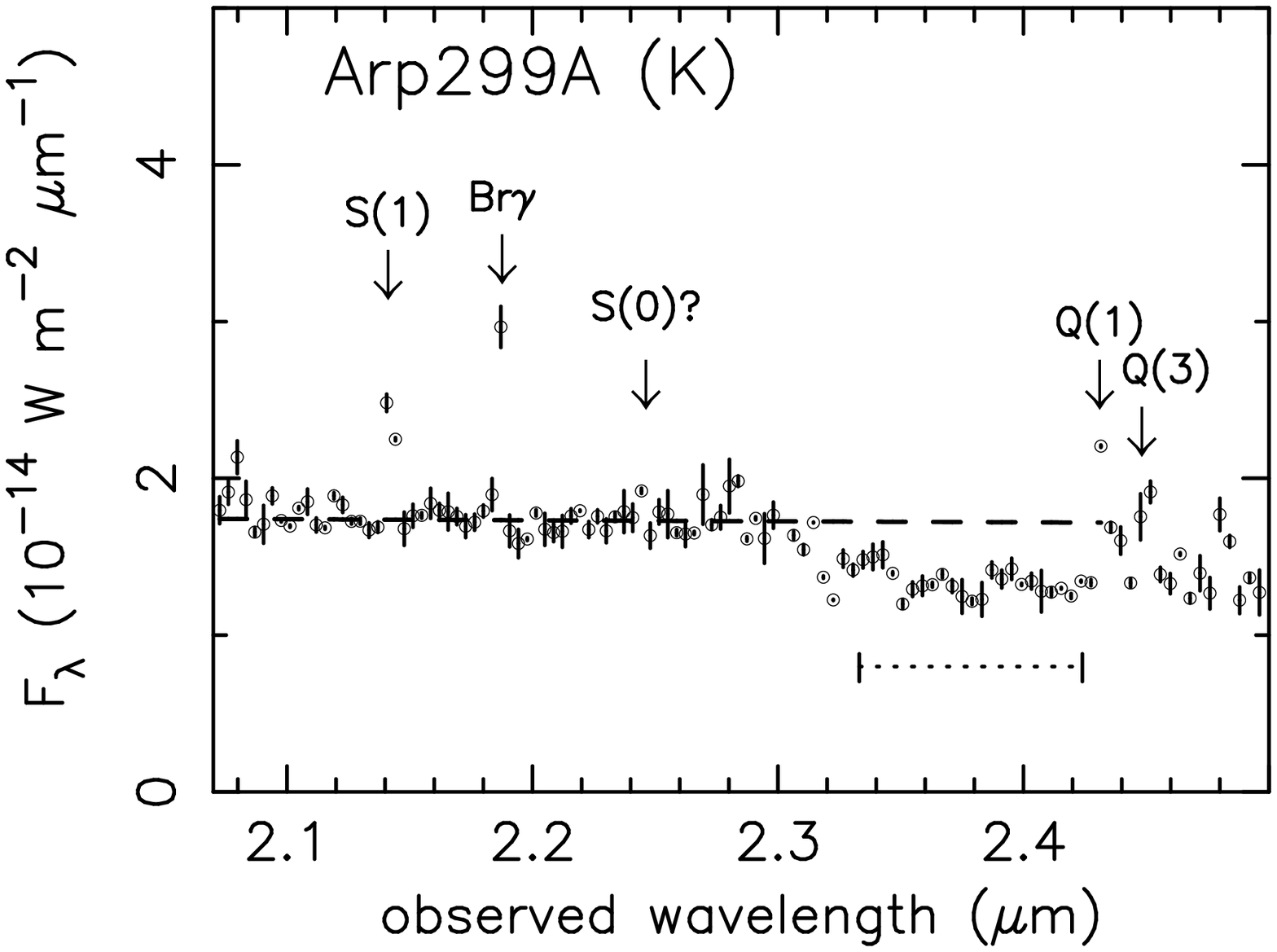}
    \FigureFile(80mm,80mm){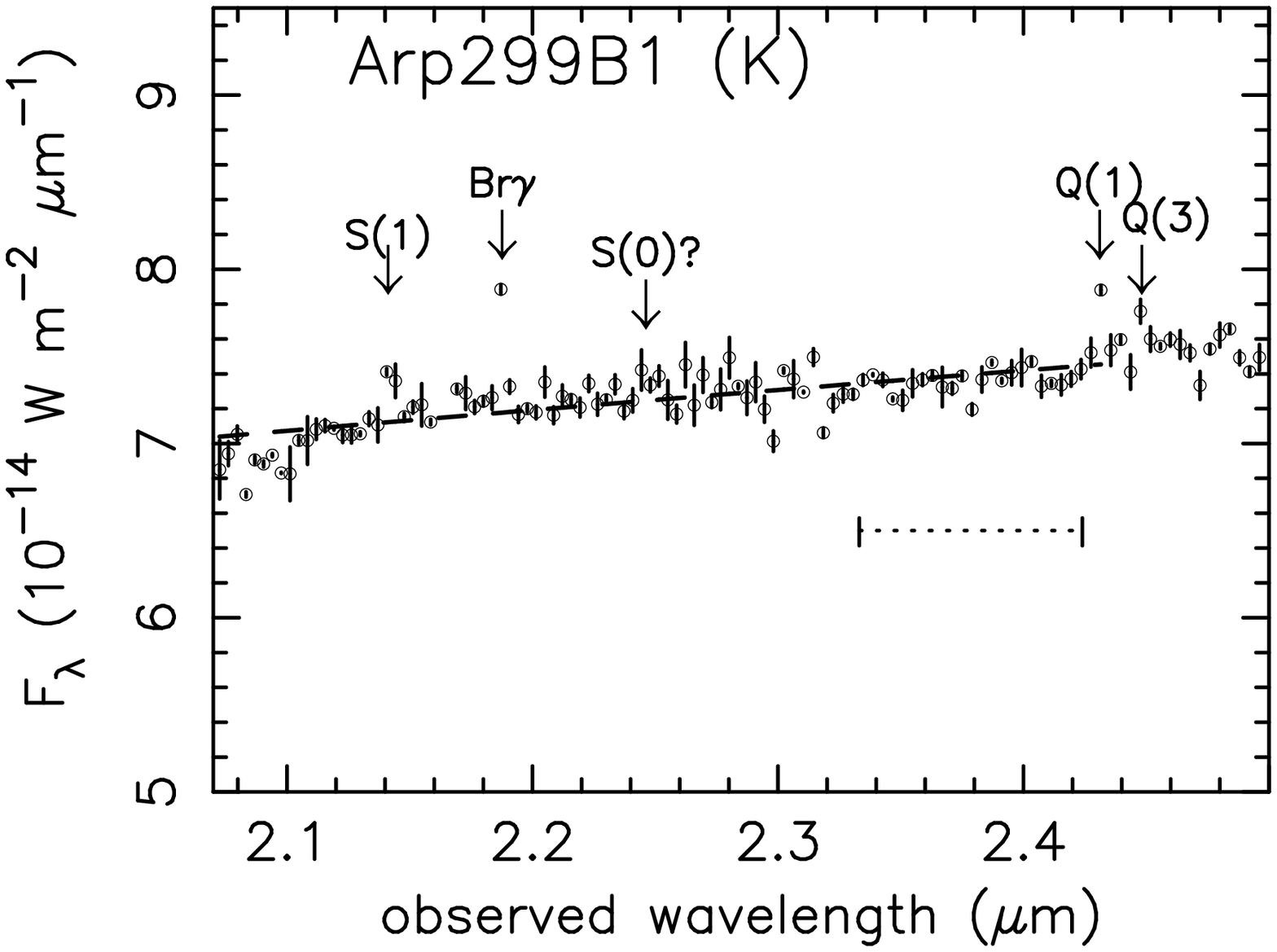} \\
    \FigureFile(80mm,80mm){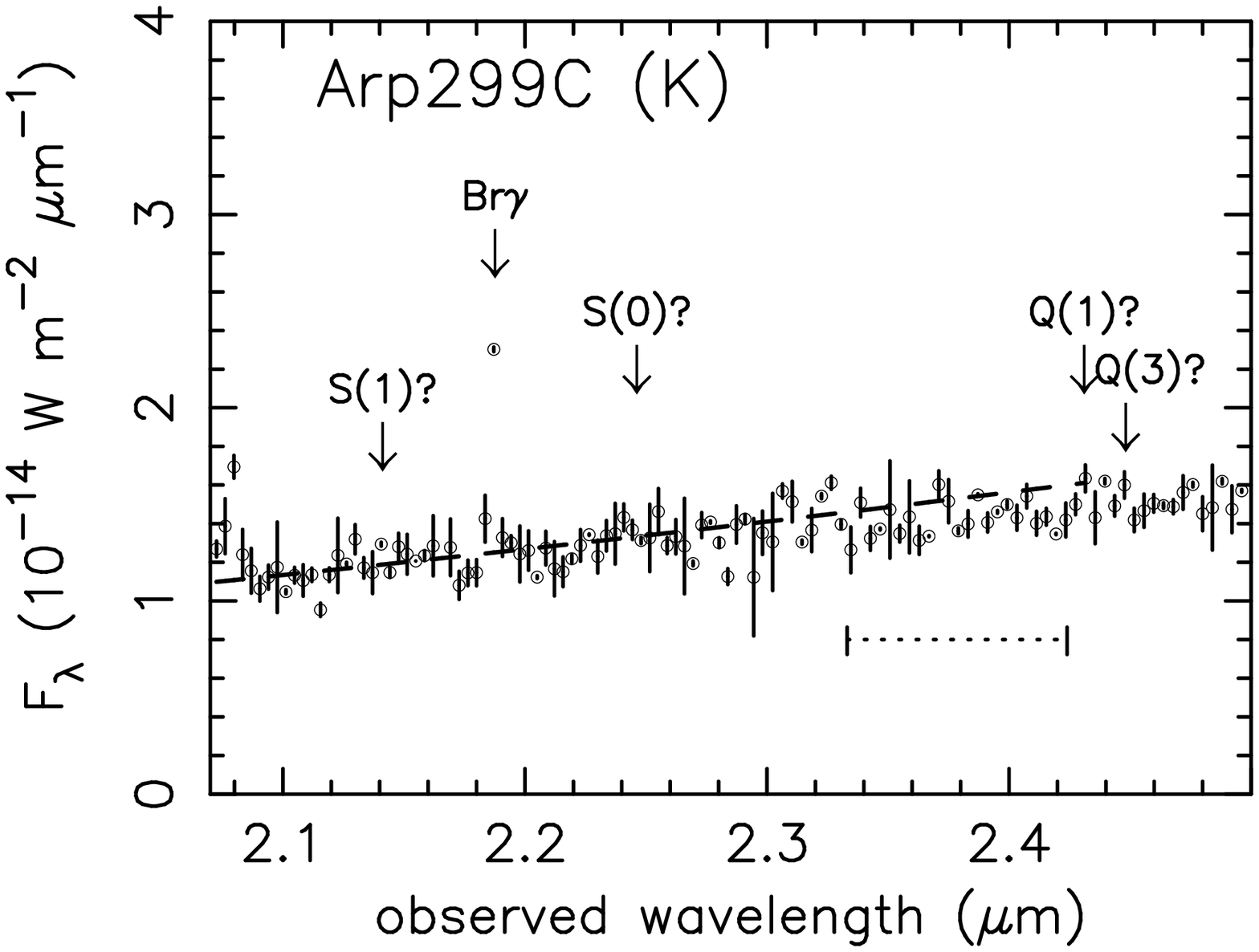}
 \end{center}
\caption{
Infrared $K$-band (2.07--2.5 $\mu$m) spectra of the Arp 299 A, B1, and
C components.
The dotted lines inserted between the two vertical lines indicate the
wavelength range of the CO absorption feature at $\lambda_{\rm rest}$
= 2.31--2.40 $\mu$m or $\lambda_{\rm obs}$ = 2.33--2.42 $\mu$m.  
The dashed lines are the continuum levels
adopted to measure the strengths of the CO absorption feature. 
Some detected emission lines are indicated.  
S(1): H$_{2}$ 1--0 S(1) at $\lambda_{\rm rest}$ = 2.122 $\mu$m.  
S(0): H$_{2}$ 1--0 S(0) at $\lambda_{\rm rest}$ = 2.223 $\mu$m.
Q(1): H$_{2}$ 1--0 Q(1) at $\lambda_{\rm rest}$ = 2.407 $\mu$m.  
Q(3): H$_{2}$ 1--0 Q(3) at $\lambda_{\rm rest}$ = 2.424 $\mu$m.  
Br$\gamma$: Br$\gamma$ emission line at 
$\lambda_{\rm rest}$ = 2.166 $\mu$m. 
}
\end{figure}

\begin{figure}
 \begin{center}
    \FigureFile(80mm,80mm){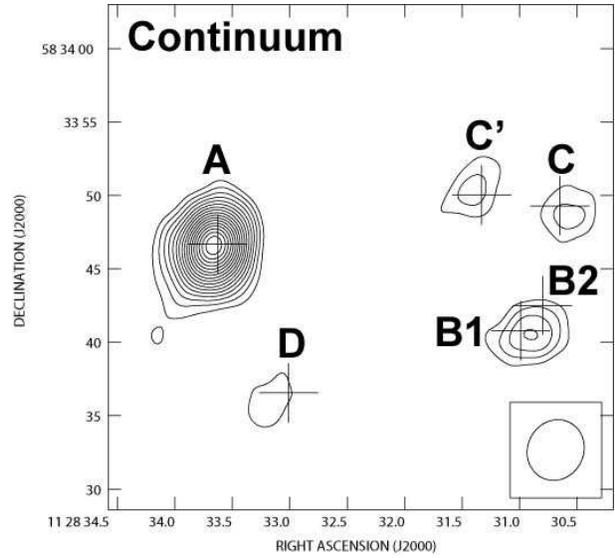}
 \end{center}
\caption{
Continuum emission map of Arp 299 at $\nu$ $\sim$ 88 GHz.
Contours start at 3 mJy beam$^{-1}$ and increase in steps of 1.2 mJy
beam$^{-1}$. 
The location of the six main components, A, B1, B2, C, C', and D are  
indicated with large crosses.
Their coordinates are taken from \citet{nef04}.
}
\end{figure}

\begin{figure}
 \begin{center}
    \FigureFile(80mm,80mm){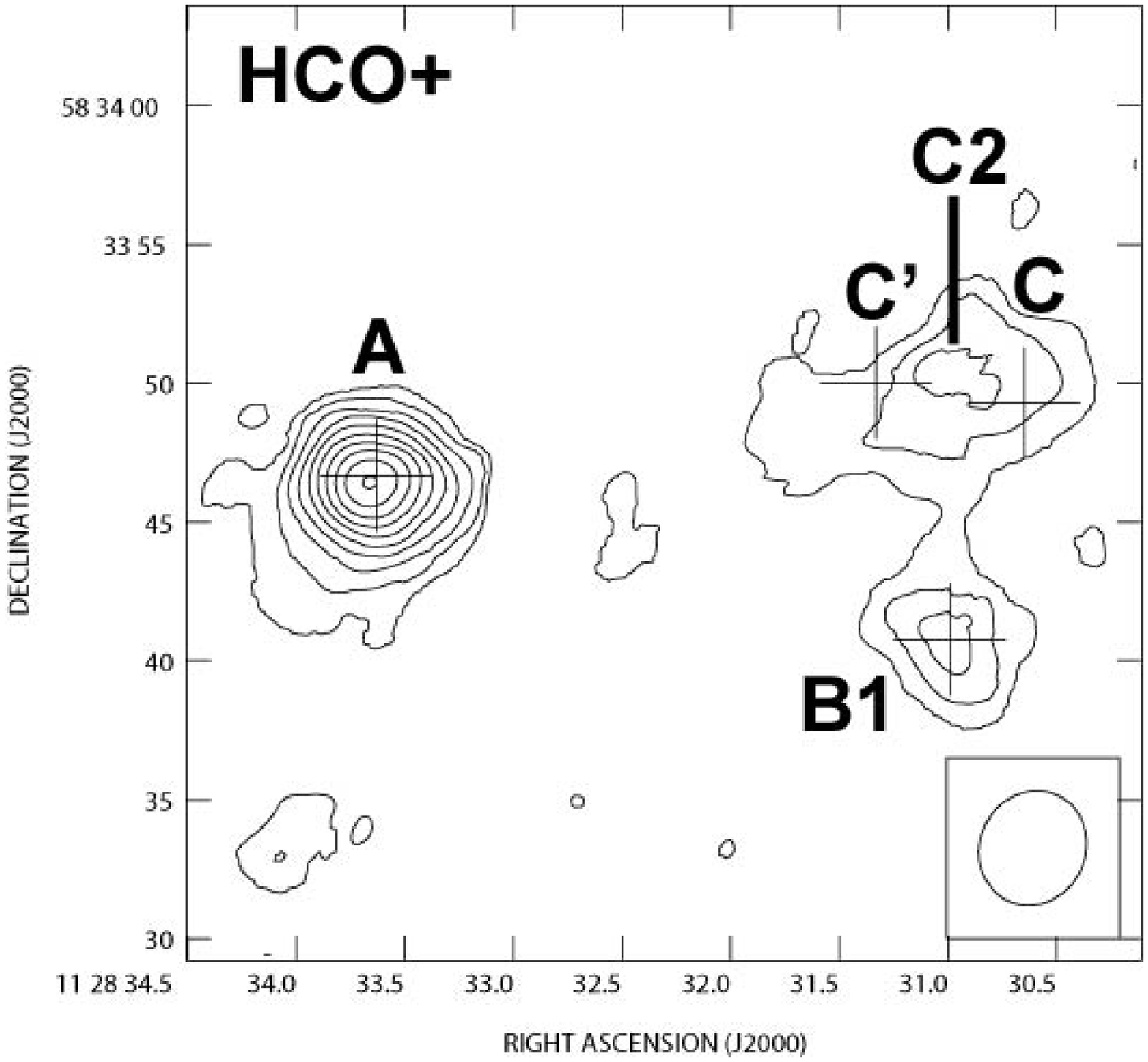}
    \FigureFile(80mm,80mm){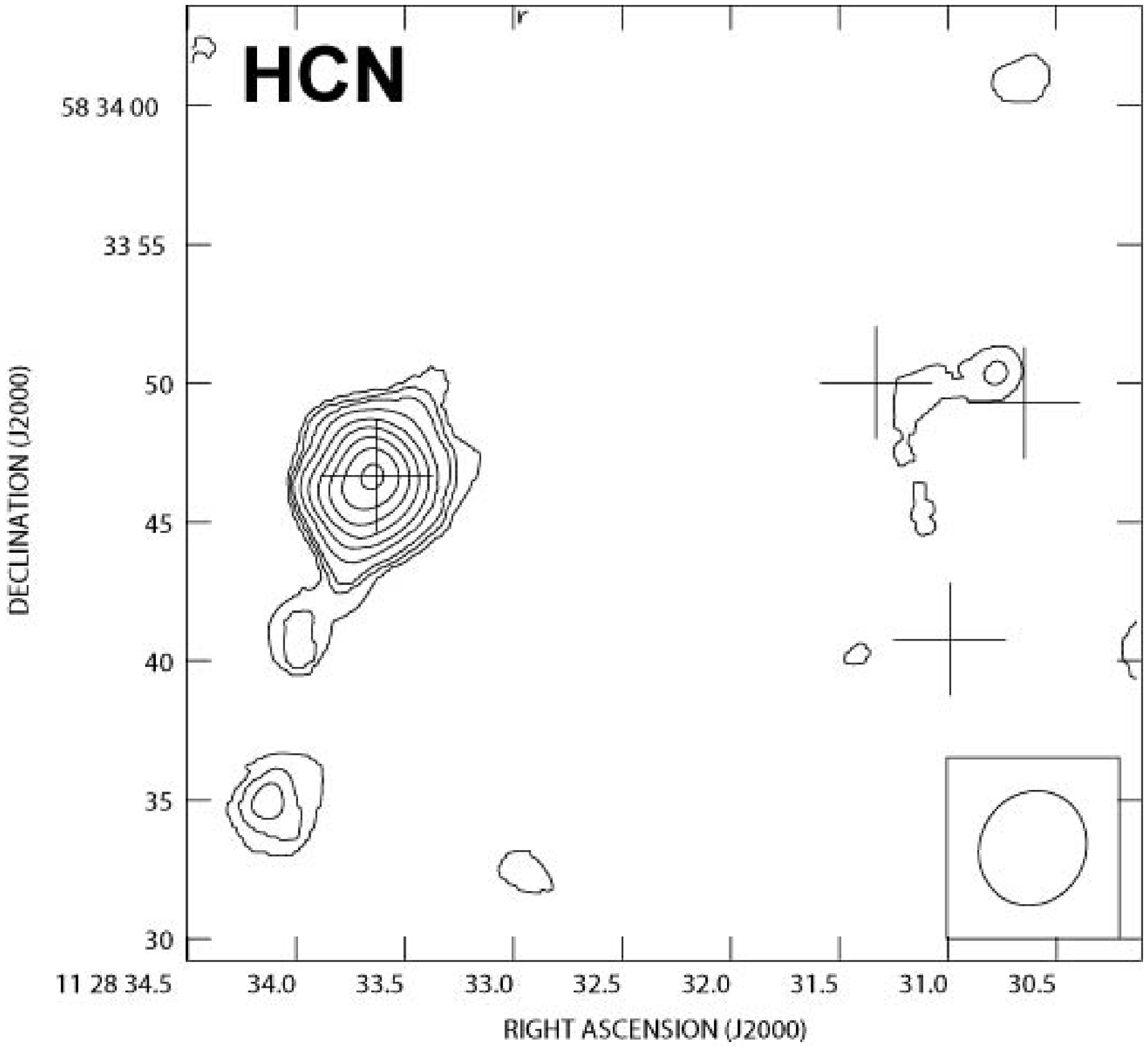}
 \end{center}
\caption{
{\it Left}: 
Integrated intensity map of HCO$^{+}$ emission.
The four major merging components (A, B1, C, C') are marked with large crosses. 
HCO$^{+}$ emission is detected between C and C'.
This region is named C2. 
Contours start at 1090 mJy beam$^{-1}$ km s$^{-1}$ and increase with
1090 mJy beam$^{-1}$ km s$^{-1}$ intervals. 
The r.m.s. noise level is $\sim$545 mJy beam$^{-1}$ km s$^{-1}$. 
{\it Right}: 
Integrated intensity map of HCN emission. 
Four large crosses denote the location of A, B1, C, and C'.
The lowest three contours are 2$\times$, 3$\times$, and 
4$\times$ 545 mJy beam$^{-1}$ km s$^{-1}$, and then increase with 
1090 mJy beam$^{-1}$ km s$^{-1}$.
The r.m.s. noise level is $\sim$545 mJy beam$^{-1}$ km s$^{-1}$. 
}
\end{figure}

\begin{figure}
 \begin{center}
    \FigureFile(80mm,80mm){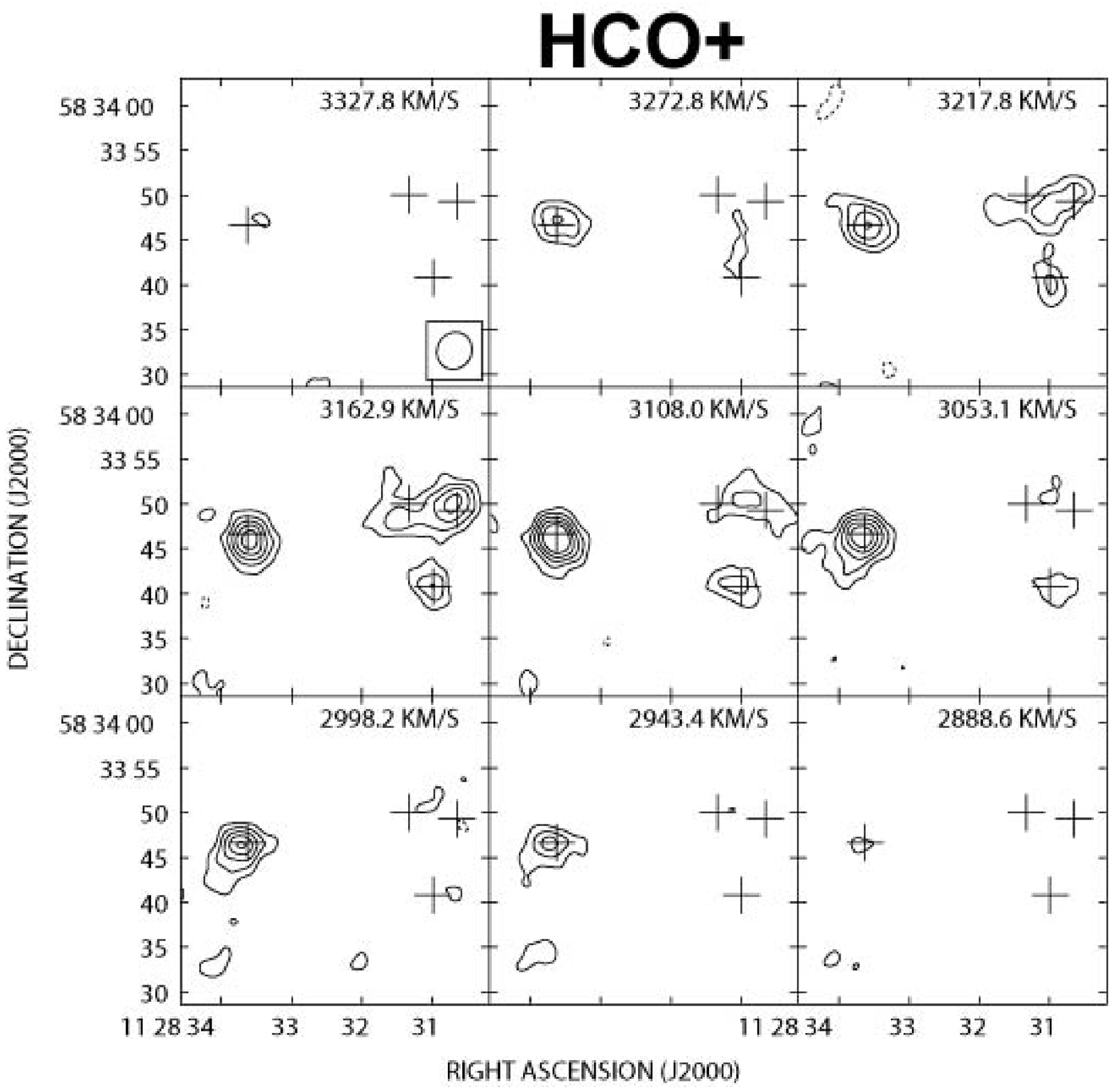}
    \FigureFile(80mm,80mm){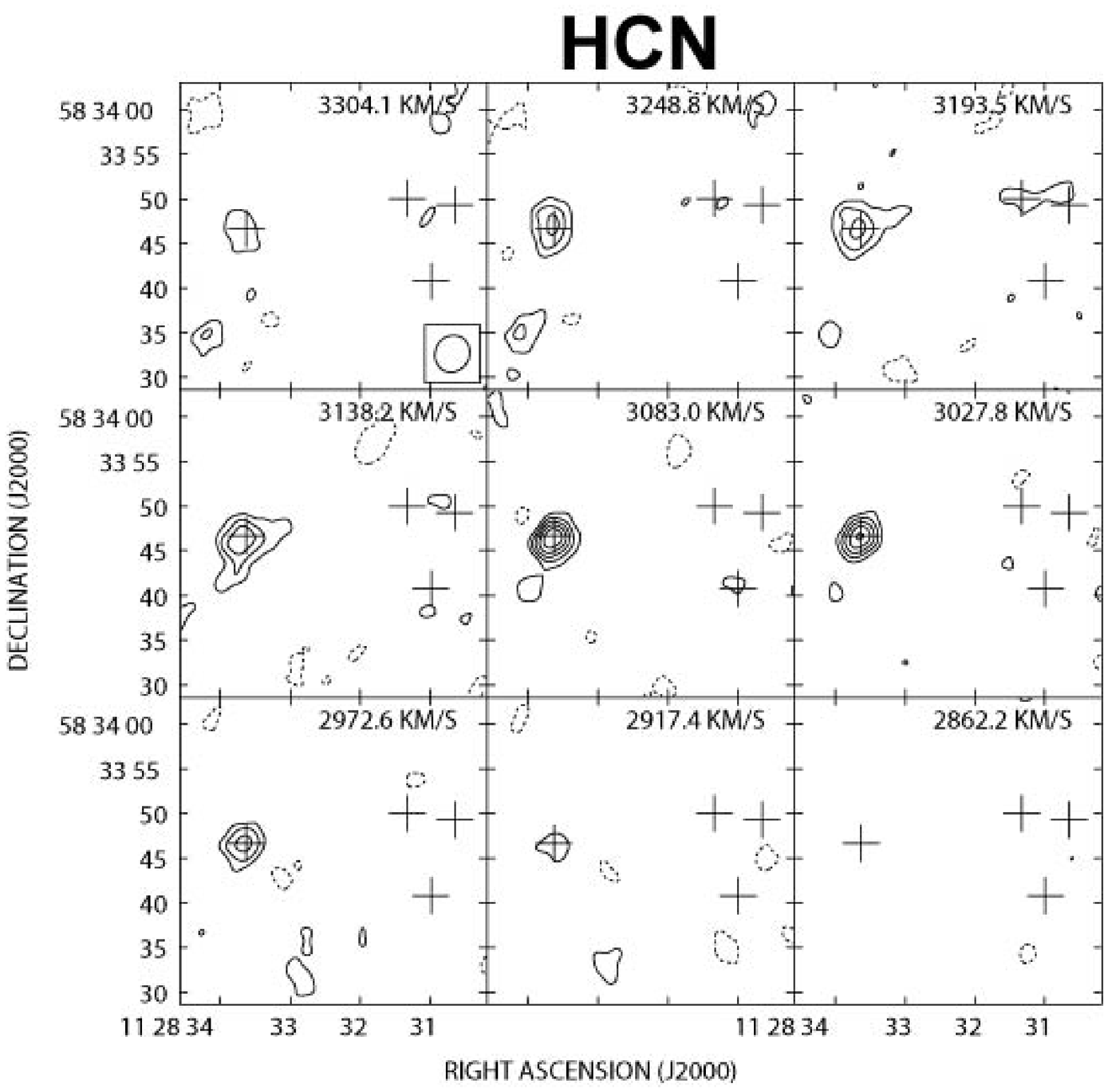}
  \end{center}
\caption{
{\it Left}: 
Channel map of HCO$^{+}$ emission in units of optical LSR velocity 
\{v$_{opt}$ $\equiv$ ($\frac{\lambda}{\lambda_0}$ $-$ 1) $\times$ c\}.
Contours start at 7.5 mJy beam$^{-1}$, and increase with 
5 mJy beam$^{-1}$ intervals.
The r.m.s. noise level is $\sim$2.5 mJy beam$^{-1}$. 
Positions of A, B1, C, and C' nuclei are plotted as crosses. 
{\it Right}: 
Channel map of HCN emission.
Contours start at 7.5 mJy beam$^{-1}$, and increase with 
5 mJy beam$^{-1}$ intervals.
The r.m.s. noise level is $\sim$2.5 mJy beam$^{-1}$. 
}
\end{figure}

\clearpage

\begin{figure}
 \begin{center}
    \FigureFile(80mm,80mm){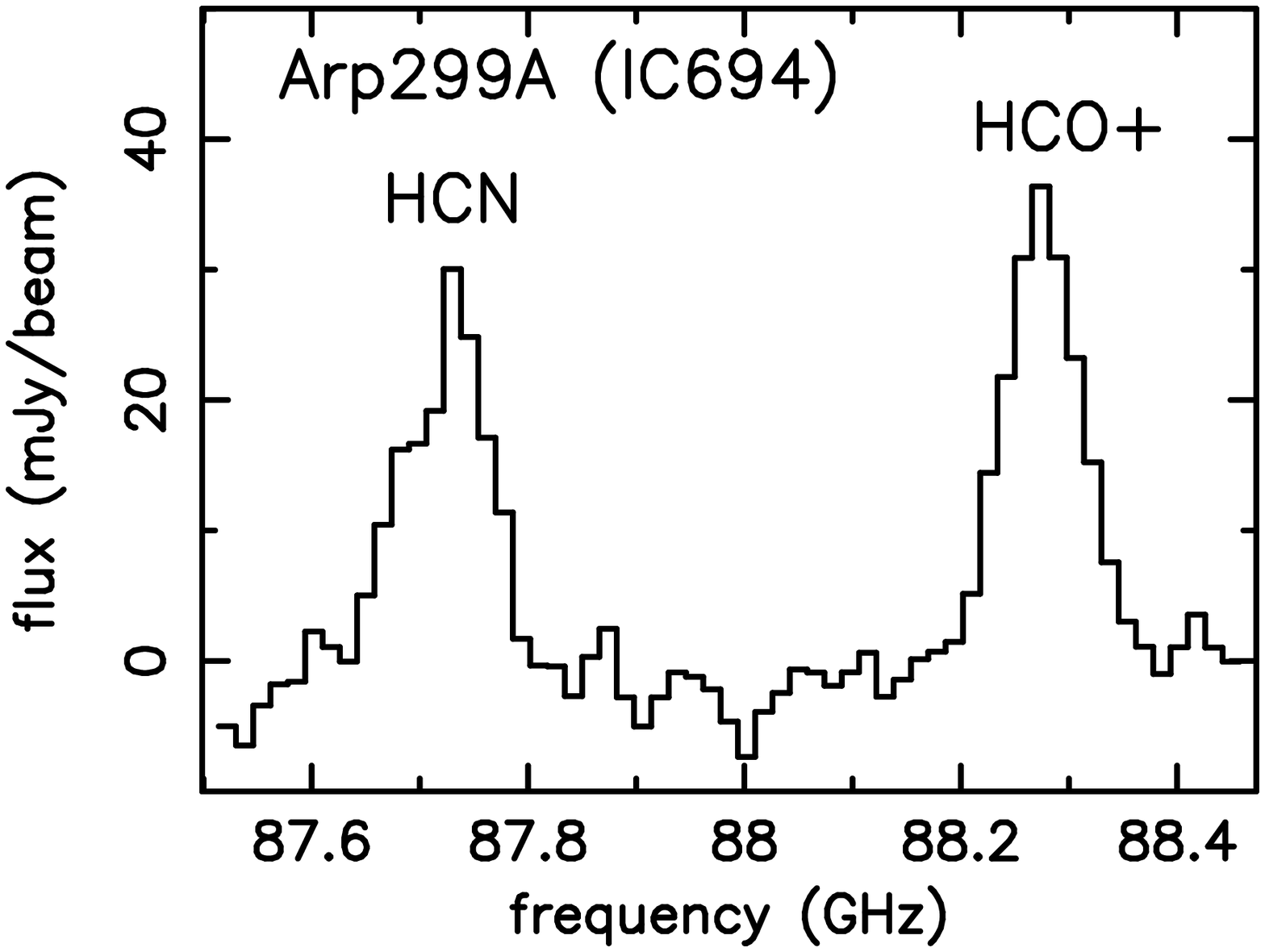}
    \FigureFile(80mm,80mm){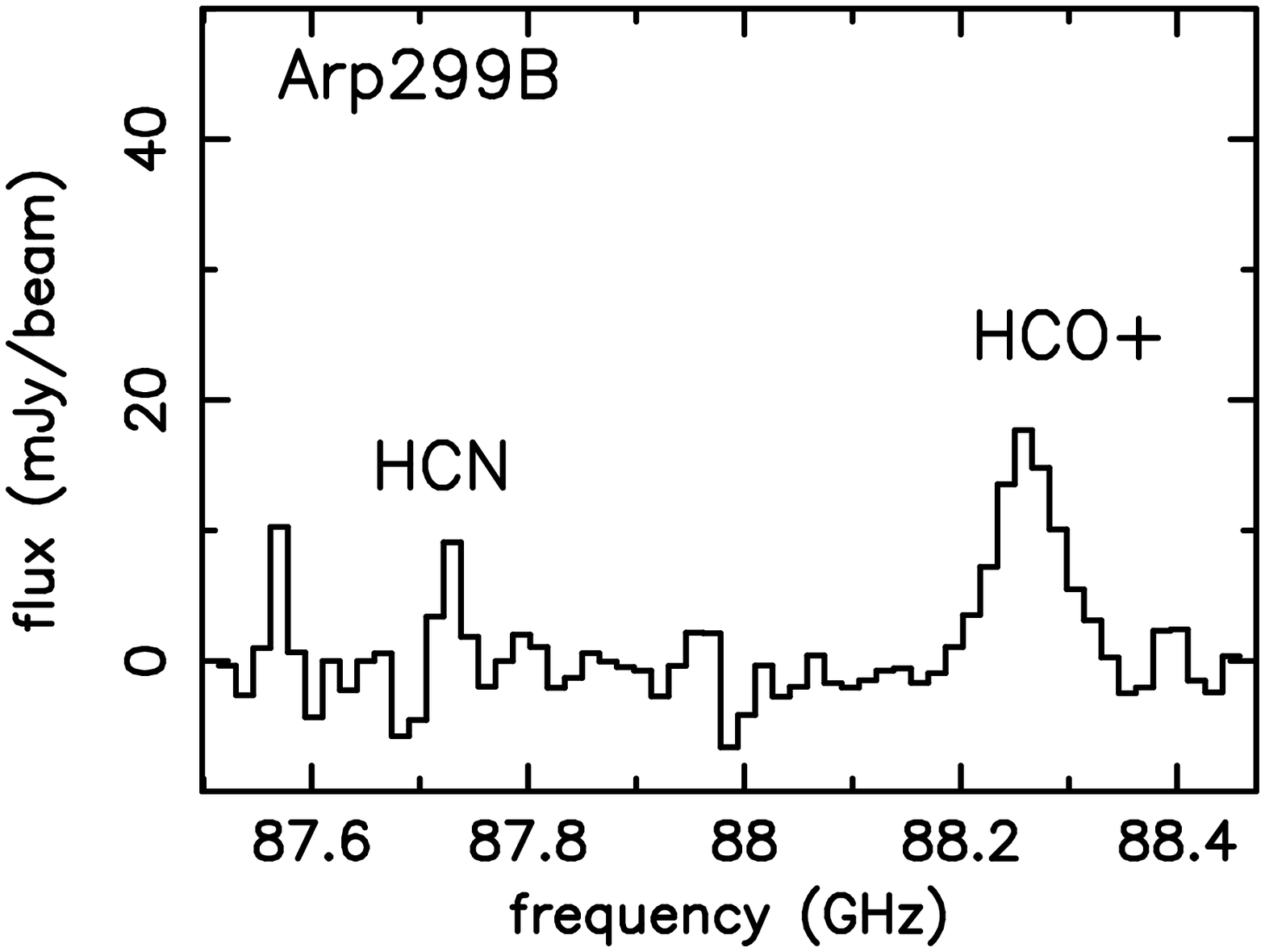} 
  \end{center}
\caption{
Spectra at the peak positions of the components A, B, C, and C2.
Peak positions are determined from our interferometric HCO$^{+}$ data, not
the coordinates provided by \citet{nef04}. 
}
\end{figure}
\begin{figure}
  \setcounter{figure}{5}
 \begin{center}
    \FigureFile(80mm,80mm){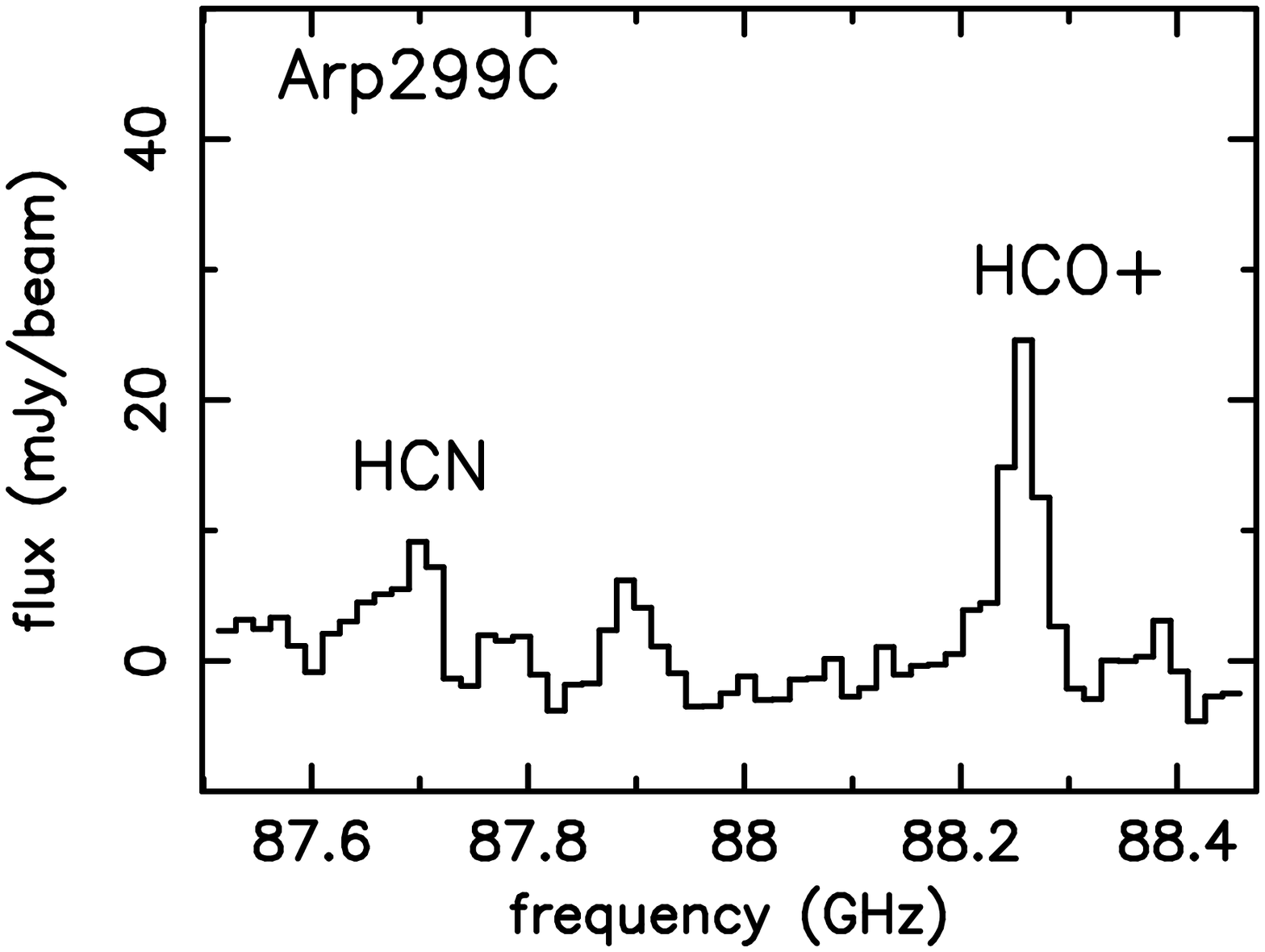}
    \FigureFile(80mm,80mm){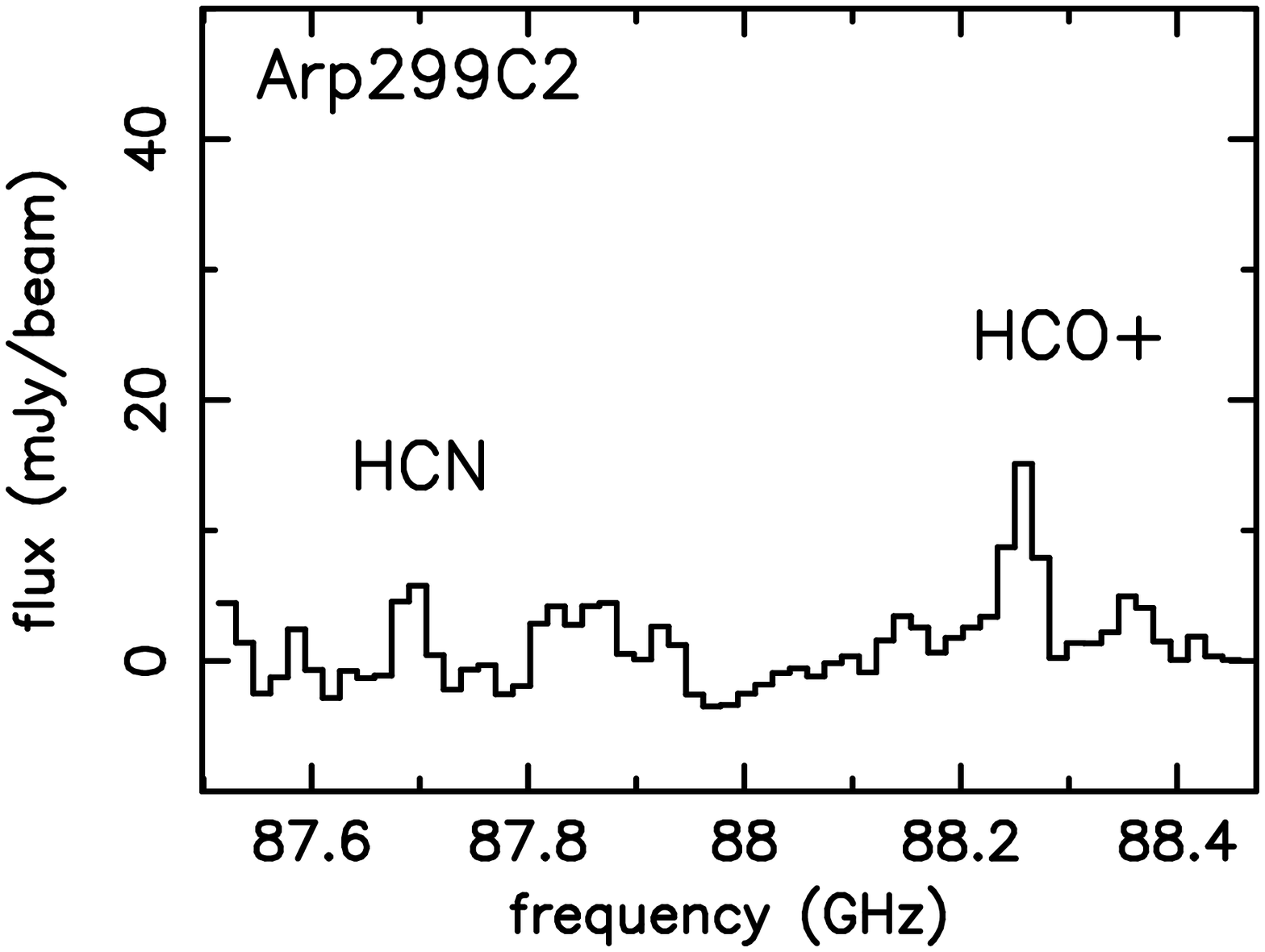} 
  \end{center}
\caption{Continued}
\end{figure}

\clearpage

\begin{figure}
 \begin{center}
    \FigureFile(70mm,70mm){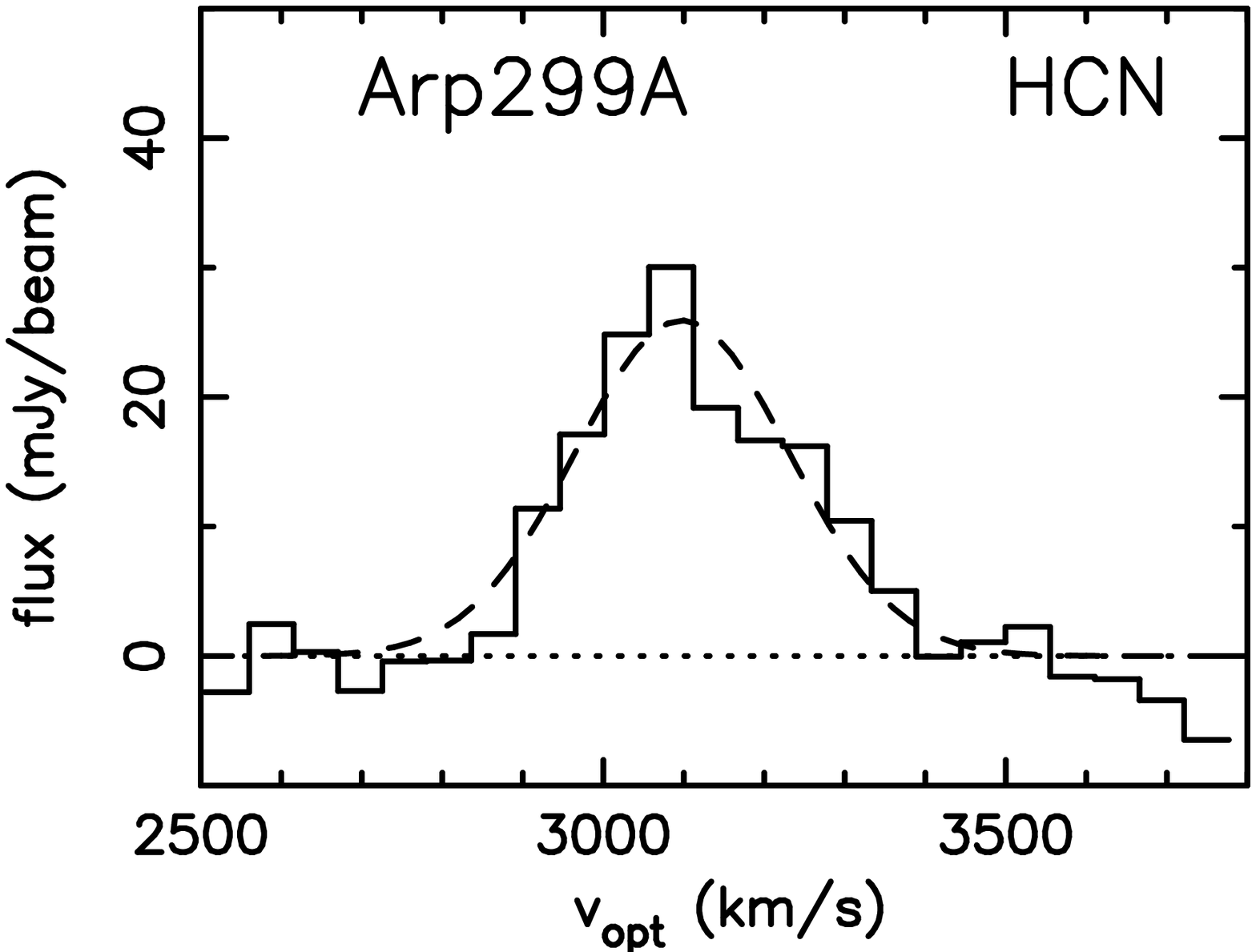}
    \FigureFile(70mm,70mm){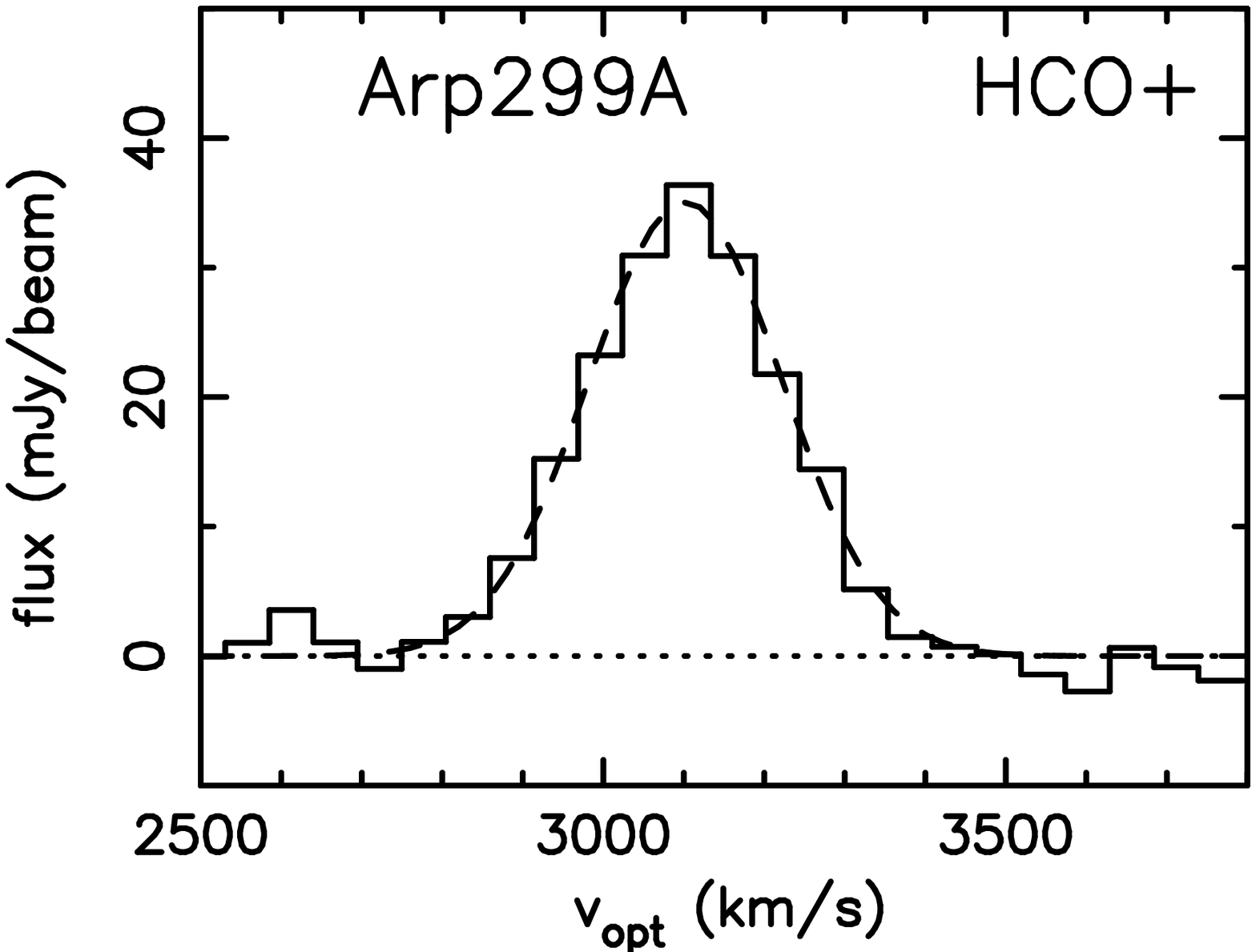} 
    \FigureFile(70mm,70mm){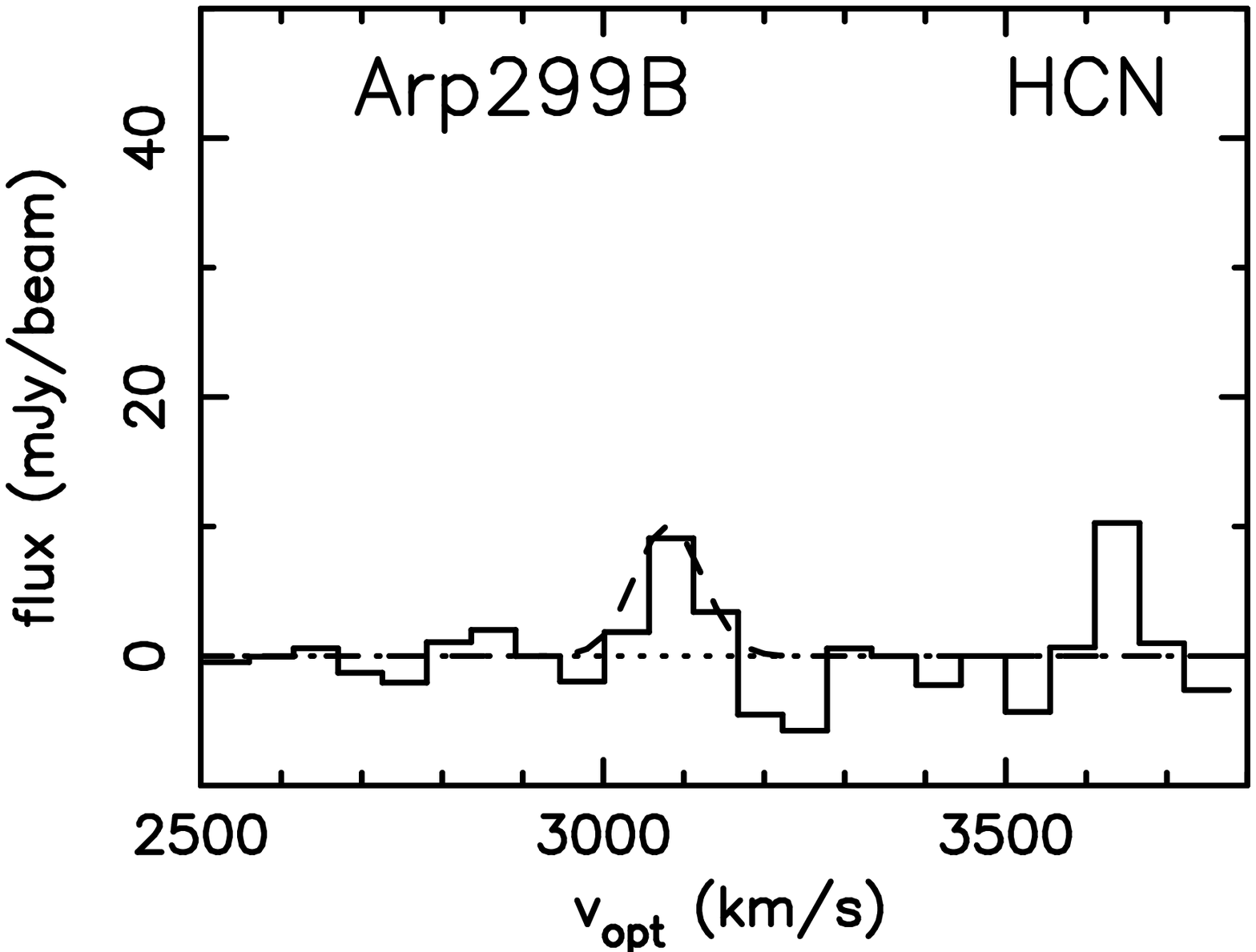}
    \FigureFile(70mm,70mm){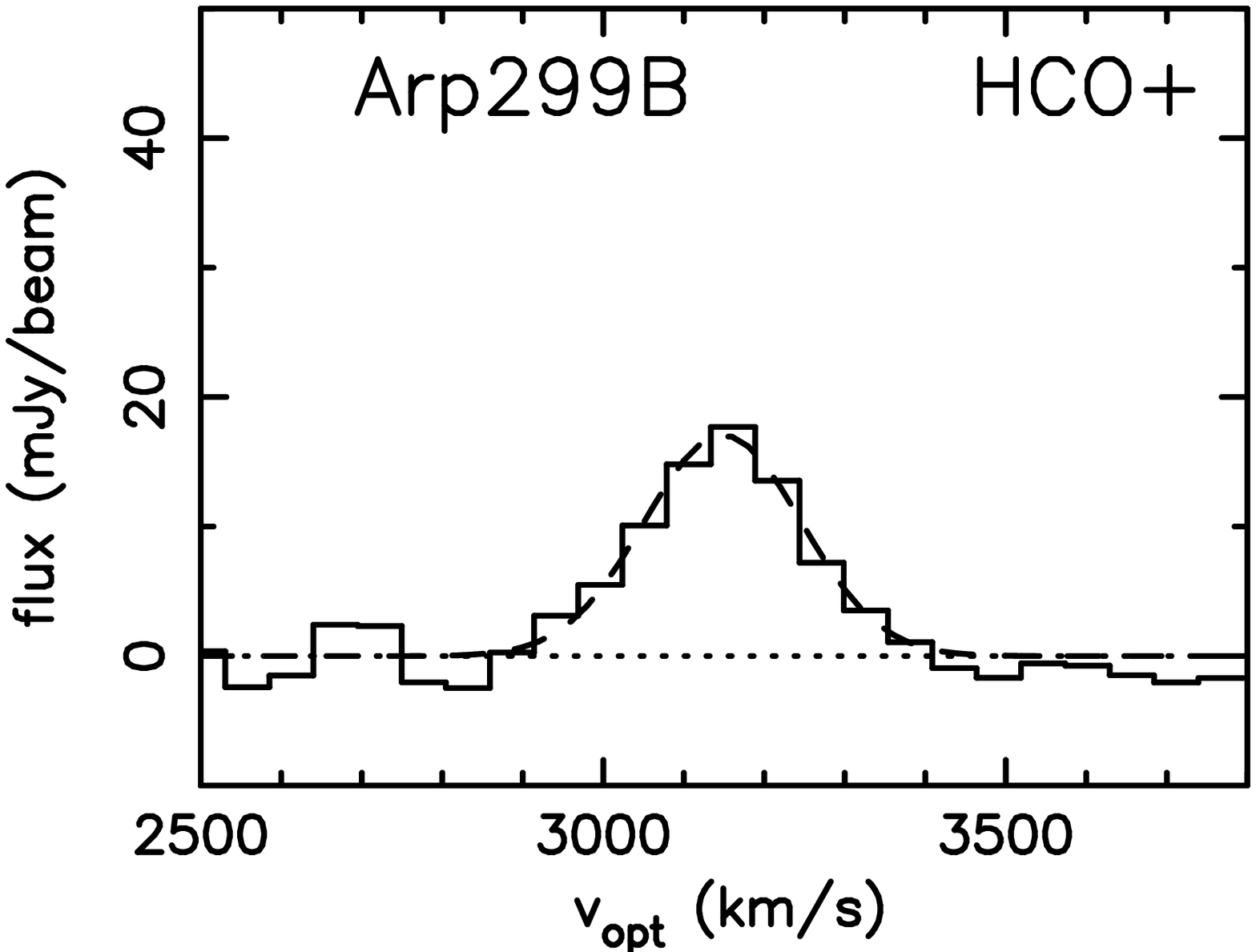} 
 \end{center}
 \caption{Gaussian fits to the HCN and HCO$^{+}$ emission lines.  The
abscissa is the LSR velocity \{v$_{\rm opt}$ $\equiv$
($\frac{\lambda}{\lambda_0}$ $-$ 1) $\times$ c\} in km s$^{-1}$ and the
ordinate is the flux in mJy beam$^{-1}$.  Single Gaussian fits are
used as defaults and are shown as dashed lines. 
Dotted lines denote the adopted continuum levels (= zero flux level).
}
 \end{figure}
\begin{figure}
 \begin{center}
  \setcounter{figure}{6}
    \FigureFile(70mm,70mm){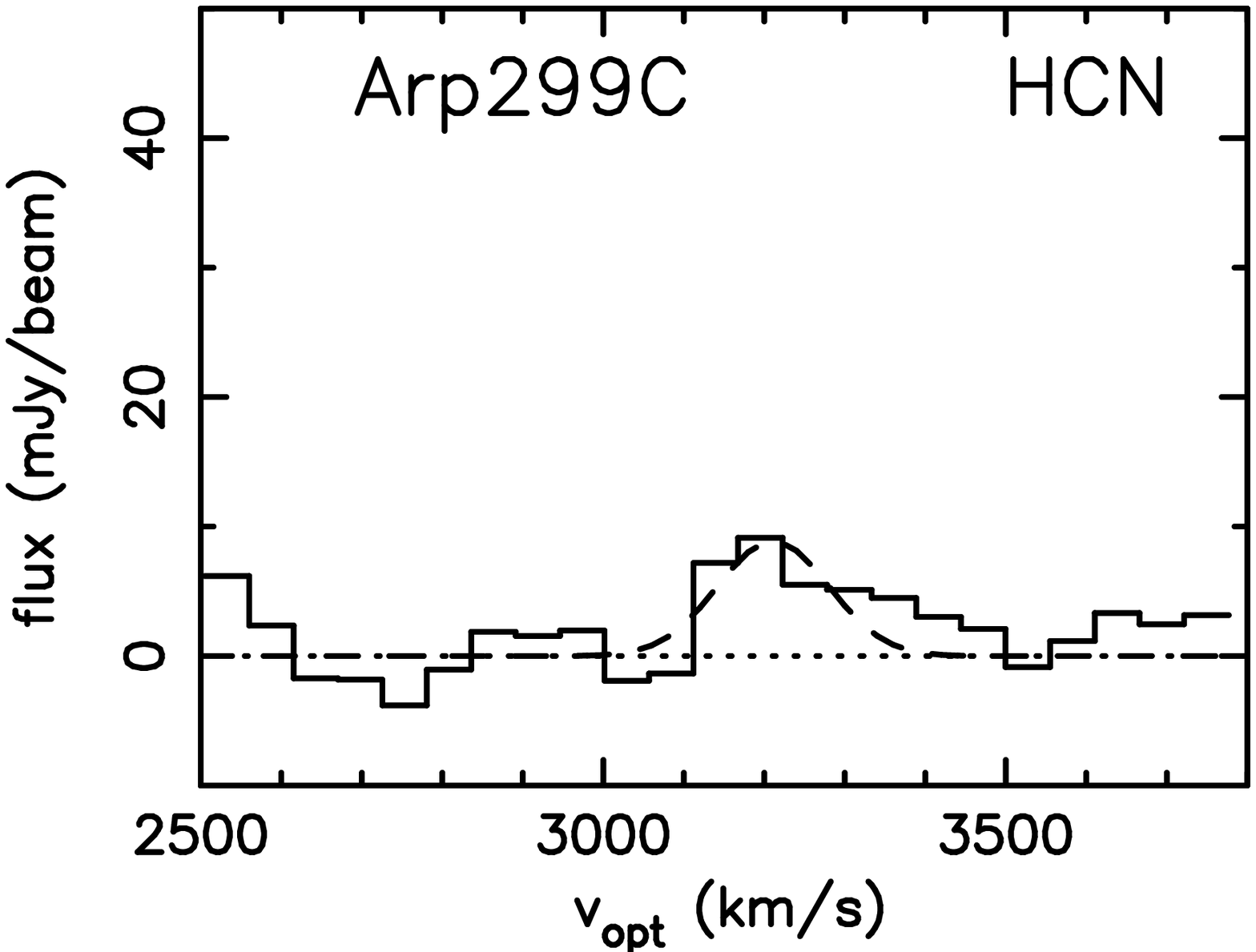}
    \FigureFile(70mm,70mm){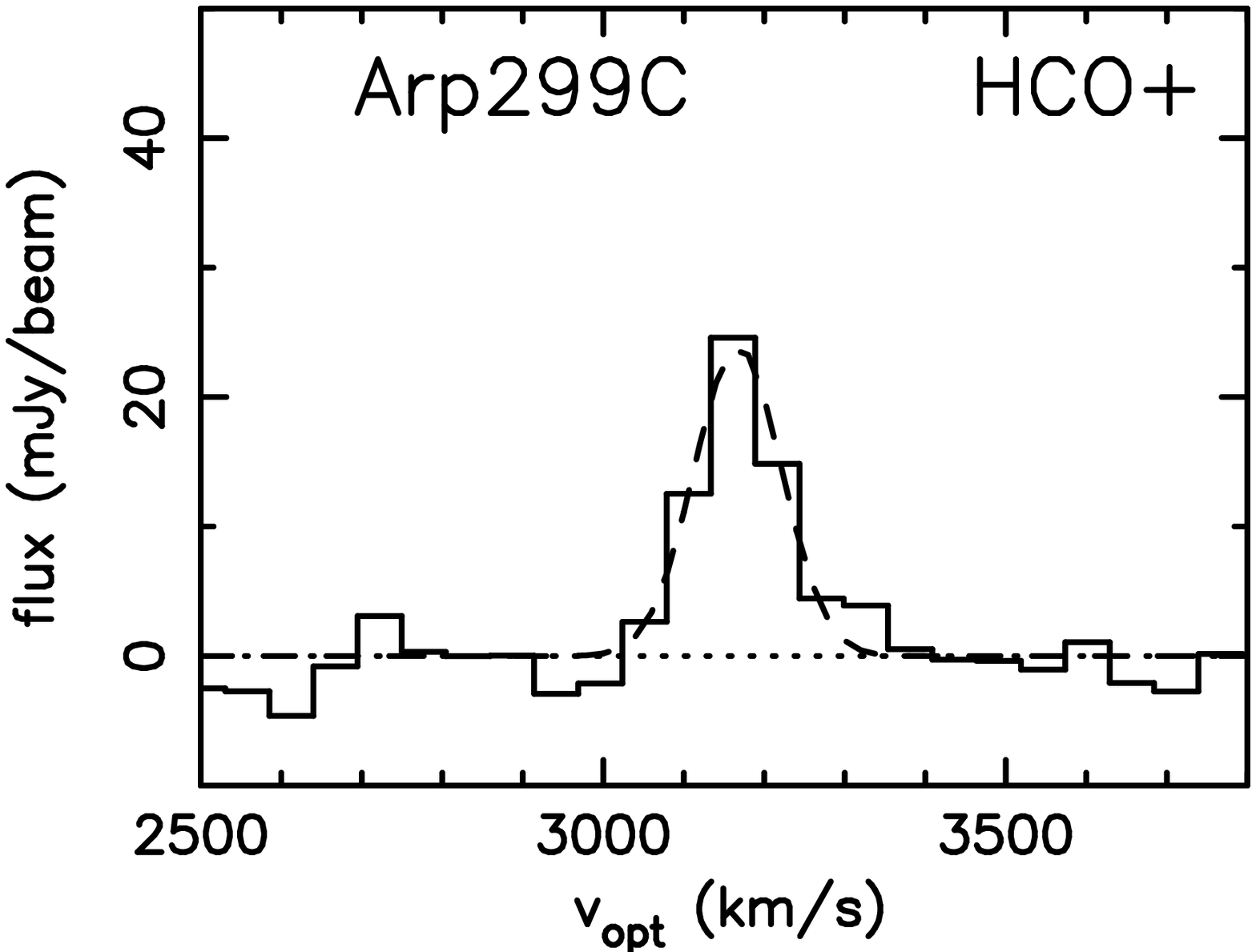} 
    \FigureFile(70mm,70mm){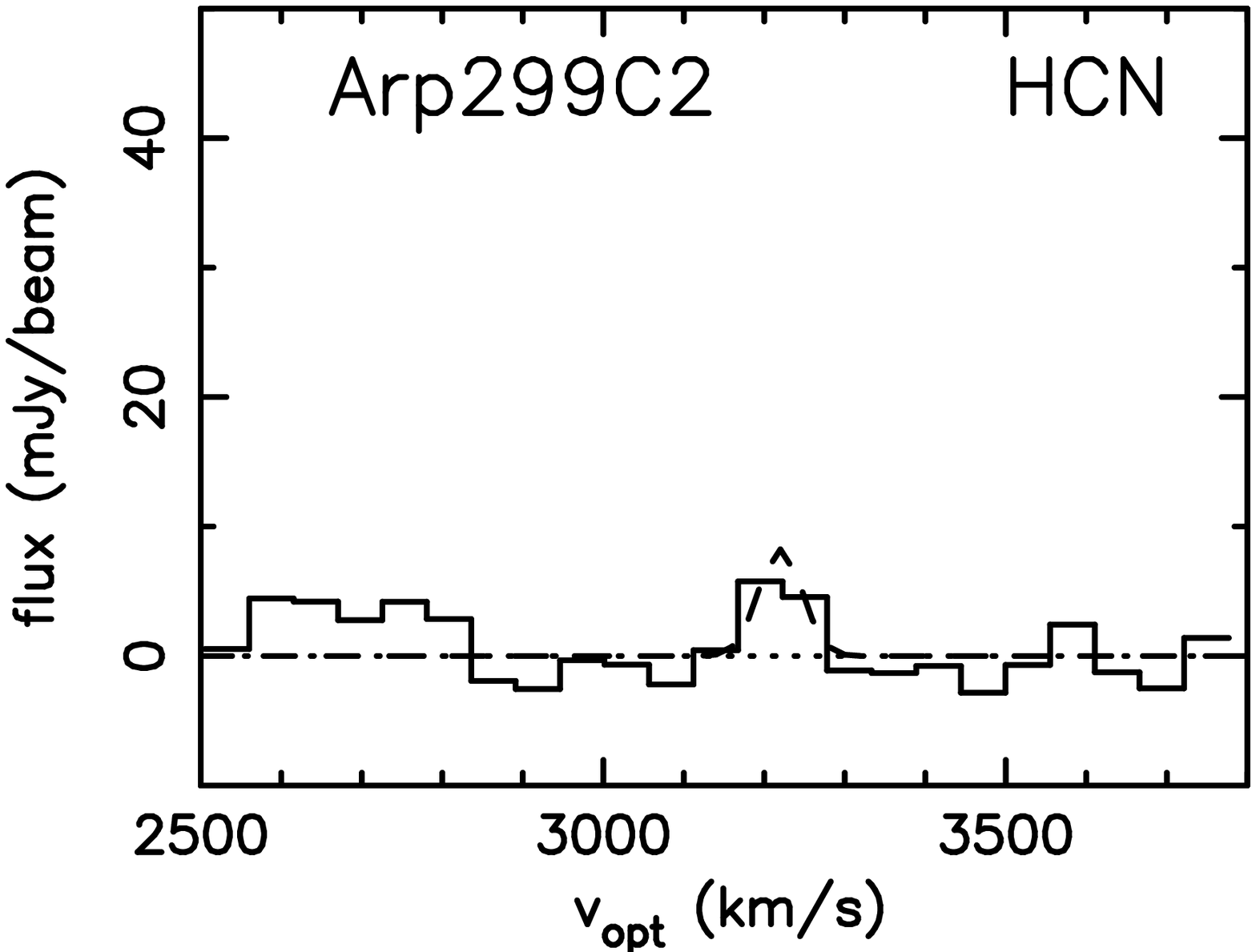}
    \FigureFile(70mm,70mm){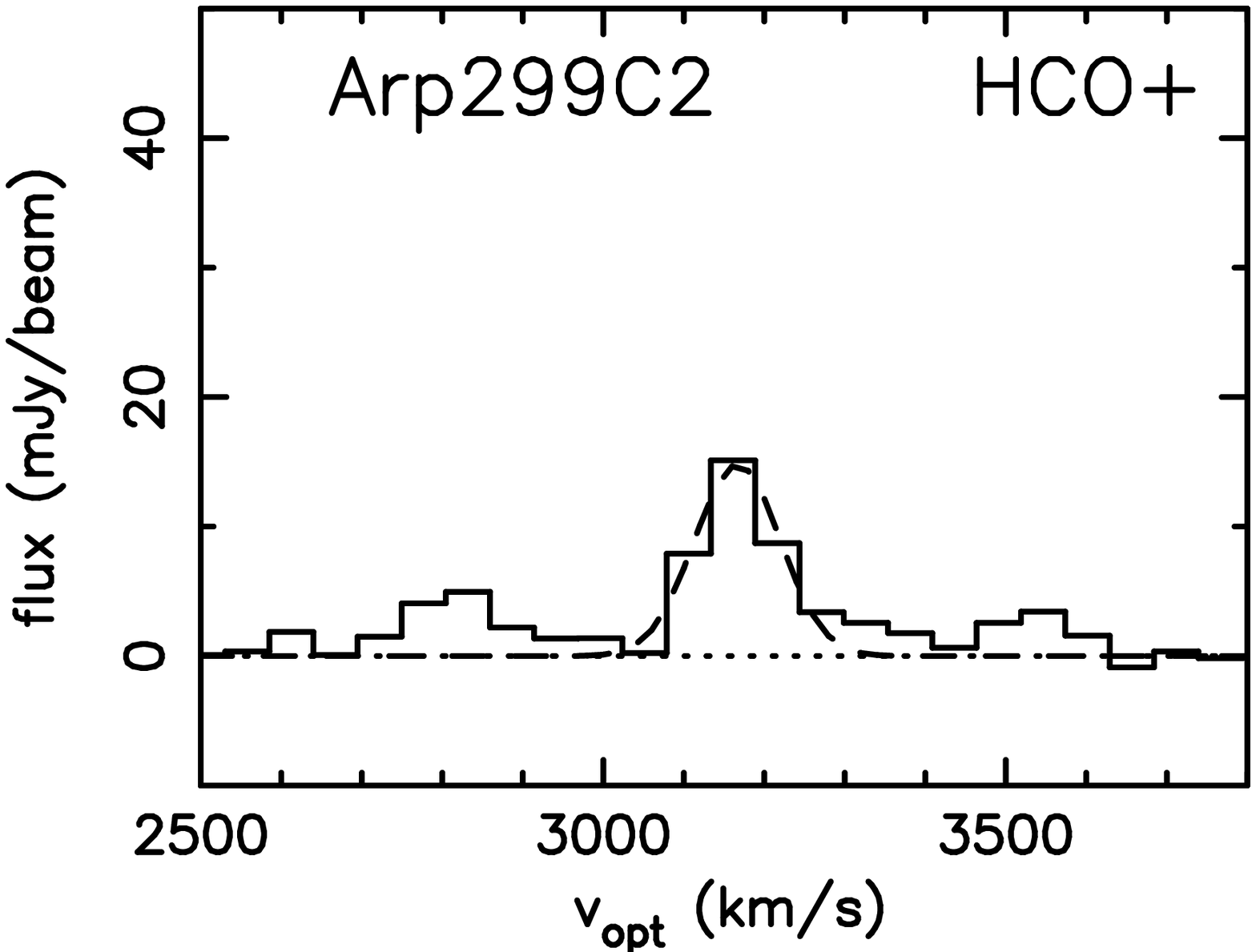}
  \end{center}
\caption{Continued
}
\end{figure}

\clearpage

\begin{figure}
 \begin{center}
    \FigureFile(160mm,160mm){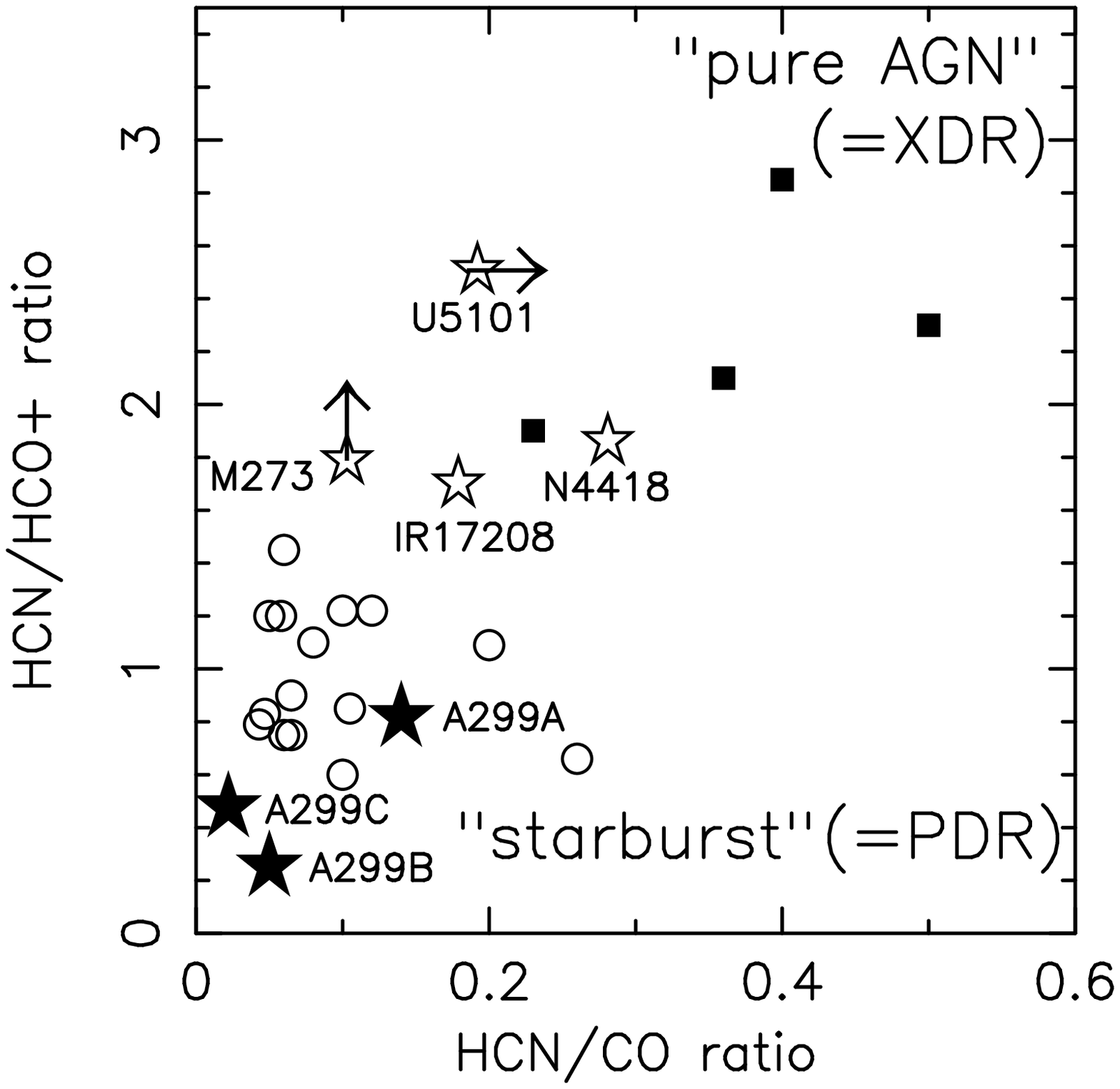}
  \end{center}
\caption{
HCN/HCO$^{+}$ (ordinate) and HCN/CO (abscissa) ratios in
brightness temperature ($\propto$ $\lambda^{2}$ $\times$ flux density).  
The three main merging components (A, B, and C) of Arp 299 are plotted
as filled stars on the lower left side.  
No data point is plotted for Arp 299 C2 because CO data used for
comparison are lacking at this exact
location.   
Other LIRGs UGC 5101, Mrk 273, IRAS 17208$-$0014, and NGC 4418
(Imanishi et al. 2004; 2006b) are plotted as open stars.   
Other data points are taken from \citet{koh01} and \citet{koh05}, where
sources with AGN-like (starburst-like) ratios are marked with filled
squares (open circles).
For all the plotted sources, HCN and HCO$^{+}$ lines were observed
simultaneously, using NMA.
The beam patterns of HCN and HCO$^{+}$ are virtually
identical for each source, and so the same nuclear regions are probed. 
HCN and CO data were not taken simultaneously for all sources,
so that their beam patterns are slightly different. 
However, this effect to our main discussions is expected to be small,
given that HCN/HCO$^{+}$ brightness-temperature ratios (not HCN/CO
ratios) are primarily used ($\S$4.3).
}
\end{figure}

\end{document}